\documentclass{mn2e} 
\usepackage{graphicx}
\usepackage{longtable}
\usepackage{supertabular,booktabs}
 \def\hs#1{\hskip#1pt}
 \def\cor{\hbox{{\it \hs{-1}C\hs{-1}oravel\/}}}
\def\cors{\cor\hs4} \def\vsini{\hbox{$v$\hs{1.5}sin\hs{2}$i$}}
\def\vsinis{\vsini\hs4} \hyphenpenalty200 \def\kms{km~s$^{-1}$}
\def\kmss{\kms\hs2} \def\:{:\hs3} 
\def\amms{\AA~mm$^{-1}$\hs3} 

\title[R Coronae Borealis]{R Coronae Borealis: Radial-velocity and other 
observations, 1950--2007}
\author[M. W. Feast et al.] {M. W. Feast$^{1,2}$, R. F. Griffin$^{3}$, G. H. Herbig$^{4}$ \thanks{Died October 2013. The Lick observations were sent by GHH to RFG and MWF in
1985. It was agreed that they should be published jointly with RFG's observations. It was
intended that GHH should comment on, and contribute to, the present text before publication, but owing to delays on the part of MWF/RFG
that was not possible. Section 2.1 is a slightly edited version of a  1986 draft by GHH.} 
and  P. A. Whitelock$^{1,2}$
\\
$^{1}$ South African Astronomical Observatory, P.O.Box 9, 7935, Observatory, South Africa\\
$^{2}$ Astronomy Department, University of Cape Town, 7701, Rondebosch, South Africa\\
$^{3}$ The Observatories, Madingley Rd, Cambridge, England\\
$^{4}$ Formerly, Lick Observatory and Institute of Astronomy, University of Hawaii }
\begin{document}
\maketitle
\begin{abstract}
  Radial-velocity observations made on more than a thousand nights are
  presented for the type star of the \hbox{R Coronae Borealis} (RCB) class.
  There are four principal sources\: the Lick Observatory (1950--1953),
  the original Cambridge radial-velocity spectrometer (1968--1991), and the
  Haute-Provence and Cambridge \cor s (1986--1998 and 1997--2007,
  respectively).  In the case of the last set the size (equivalent width)
  and width (expressed as if \vsini ) of the \cors cross-correlation (`dip') profiles are
  also given, and the variation and complexity of those profiles are discussed.  Although there is often
  evidence of cyclical behaviour in radial velocity, no coherent periodicity is found in any of
  the series. From time to time, and especially over 100 days before the great decline of
  2007, the atmosphere was highly disturbed, with evidence of high-velocity components.
  We suggest that those are associated with large turbulent elements and result in
  mass ejection to sufficient distances for the formation of soot and other solids and thus    
  the initiation of RCB-type declines.  We associate the changes in light and radial velocity
  near maximum light primarily with the combined effect of such turbulent elements, and not with coherent pulsation.  There is some evidence for a variation in the mean radial velocity on a time scale of about ten thousand days.

\end{abstract}
\section{Introduction}

R Coronae Borealis is the type star of a small group of variables (the RCB stars) which,
despite work over a considerable number of years, remain something of a
puzzle.  There are two main, related, aspects to this puzzle -- first, their
physical nature and behaviour, and secondly, their place in stellar
evolution. The group has been reviewed on a number of occasions
(e.g. Clayton 1996, Feast 1996).

The stars are carbon-rich and hydrogen-poor and have a range of anomalies in
other elements (see for instance the review of Rao \& Lambert 1994).  They
also have infrared excesses due to radiation from a circumstellar shell with
a temperature of order 1000\hs1K.  At random intervals they undergo dramatic
declines in light (up to 8 mag in $V$) (RCB events), which are believed to
be due to the formation of carbon, or carbonaceous, particles (`soot'). Those particles
cannot be in the form of a uniform shell around the star, since the increase
in infrared flux that would then be expected is not observed (Forrest et
al. 1971).  The events are attributed (Wdowiak 1975) to puffs of dust
 that form in material ejected from the
star in the line of sight.  Such puffs, ejected from time to time in random
directions, feed the dust shell seen in the infrared.  Infrared studies of
the RCB star RY~Sgr, which pulsates with a considerable amplitude, show
that overall the star itself is not significantly affected during an event
(Feast 1979).  That indicates that any surface disturbance related to an RCB
event must be confined to a small area.  Other evidence which requires, or
is consistent with, the puff model is summarized in Feast (1986).  More
recently, direct support for such a model for RCB stars has come from
high-resolution direct imaging, interferometry, and, in the case of UW~Cen,
changes in the illumination of a circumstellar reflection nebula (Jeffers et
al. 2012, Bright et al. 2011, Clayton et~al. 1999).

In the case of R~CrB itself there is also a large ($\sim8$ pc) cool shell
which is possibly the debris from the ejection of the star's hydrogen-rich
envelope a few times $10^4$ years ago (Gillett et al. 1986); see also 
Montiel et al. (2015).

RY Sgr is unusual in showing clear coherent pulsations with a period of about
38 days. Most other RCB stars show smaller-scale variations in light and radial
velocity on a time scale of the order of a month.  In the case of R~CrB
these are $\sim$0.2 mag and $\sim$10 \kms.  It has often been suggested
that all the RCB stars are radially pulsating variables and that
the formation of the dust puffs is associated in some way with their
pulsation, though that view has been questioned (e.g. Feast 1996).

Once dust is formed it will be blown away from the star by radiation
pressure, dragging gas with it, and outward velocities of $\sim$200 \kmss
or more
are observed at that stage (see reviews cited above).  It~is not clear,
however, how low-amplitude pulsations can raise gas to regions cool enough
for dust to form, and that has remained a concern.  

There are two competing basic models for the RCB stars: either they are
`born again' AGB stars (i.e. post-AGB stars undergoing a final helium shell
flash, e.g. Fujimoto 1977, Renzini 1979) or else the result of mergers of CO
white dwarfs with helium ones (Webbink 1984).

The purpose of the present paper is to present several long-term
radial-velocity and other studies of R~CrB itself and to make a preliminary
analysis of those data.  In particular, we investigate the nature of
small-scale velocity and other variations of the star and their possible
relationship to dust ejection, as well as phenomena occurring at the onset
of major declines in light.

\section{Observations}
\subsection{The Herbig (Lick) Series}\label{Sec_herbig}

\begin{figure}
\vspace*{-4.5 cm}
\begin{center}
 \includegraphics[width=3.4in]{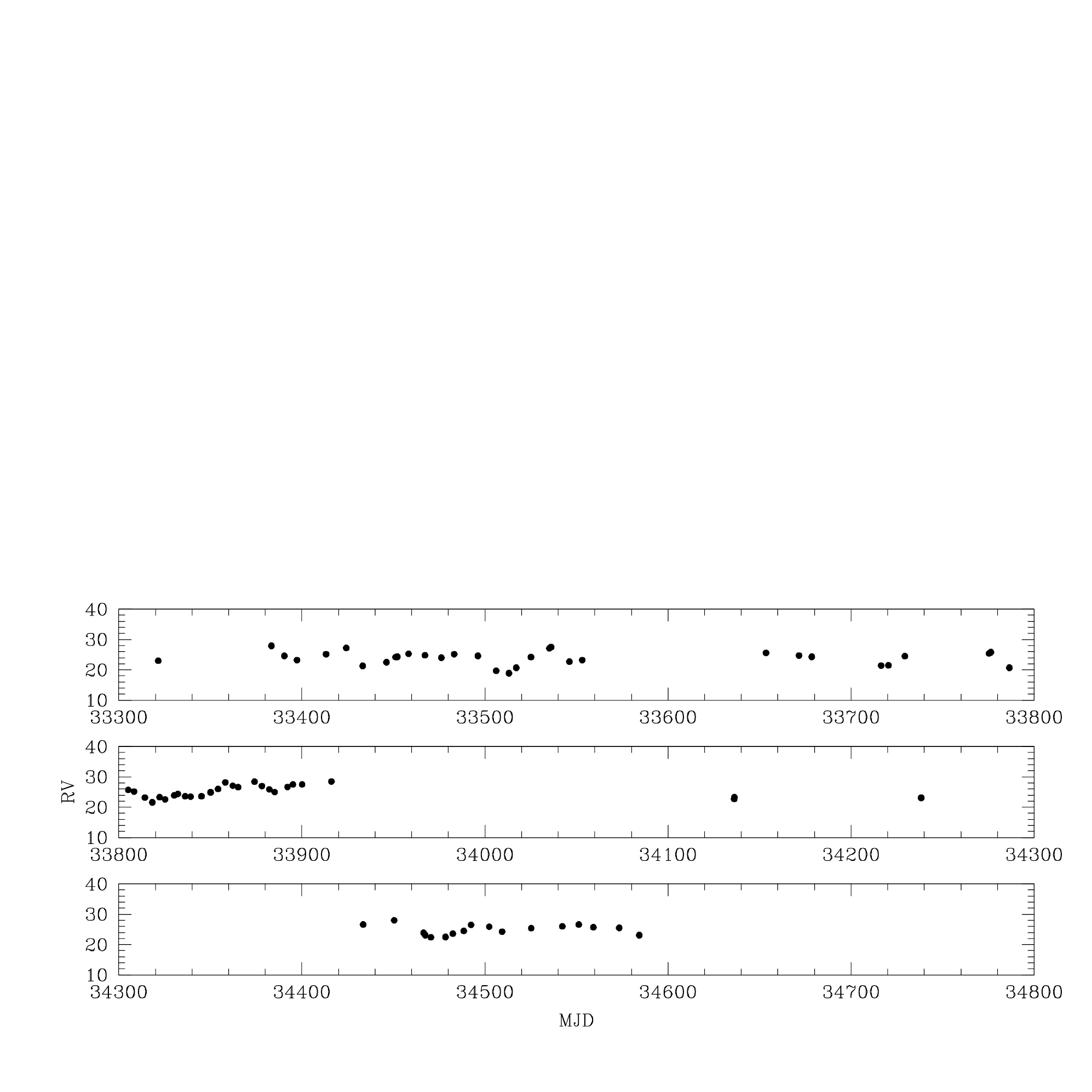} 
\vspace*{-0.5 cm}
 \caption{The radial velocities (RV) from Lick Observatory as a function of Modified Julian Date (MJD).}
   \label{herbig}
\end{center}
\end{figure}
 
R~CrB was observed with the 3-prism Mills spectrograph attached to the Lick 36-inch refractor at almost every opportunity during the observing seasons of 1950, 1951 and 1953.  At those times, the
star was near normal brightness (see section 3.4.1), and 73 acceptable spectrograms
(dispersion 11 \amms at $\lambda$4500 \AA) were obtained, with one further observation in 1956. The spectra were
measured in the 1950s with a Hartmann spectrocomparator (Hartmann 1906,
Moore 1935) against a standard plate of a reference star, but the results
were not published. They were re-measured in the 1980s by an automated
procedure that is equivalent to the Hartmann method and also to a
radial-velocity spectrometer.  A~transmission scan of an R~CrB spectrogram
was digitized at 2.5-$\mu$m intervals by the computer-controlled
Lick/Gaertner two co-ordinate microphotometer and was then cross-correlated
with a reference scan of a standard R~CrB plate.

The differential velocities thus obtained were reduced to the Lick or IAU
velocity system by measuring in the same way, against the same R~CrB
standard, a number of Mills plates of standard stars (Pearce 1957) taken
during the R~CrB programme.  The stars measured, together with the adopted
velocity and the number of plates measured, were: $\gamma$~Cyg \hbox{($-5.4$
\kms, 7)}, \hbox{$\alpha$~Per ($-2.3$, 7)}, and $\beta$~Aqr \hbox{(+6.7,
5)}. The scatter of the measured velocities of those standards about the
adopted values, after application of the same overall zero-point correction
to the R~CrB reference as was applied to the velocities of R~CrB itself,
correspond to a standard deviation (for an individual Mills velocity) of
$\pm$0.3~\kms, which is probably optimistic. The internal agreement of the
individual sections of each plate (the spectrograms were measured in
20-\AA\hs3 sections) yields a probably more realistic value of 
$\pm$0.6~\kms.  The (heliocentric) radial velocities (\kmss) of R~CrB so obtained 
are given in the Appendix Table A1 together with the Modified Julian Date (MJD). 
The radial velocities are plotted against MJD in Fig.~1.

\subsection{Radial velocities from spectrometers}

About 50 years ago one of the present authors developed the method of
cross-correlation, now in universal use, for measuring radial velocities.
The prototype instrument (Griffin 1967) operated at the coud\'e focus of the
Cambridge 36-inch reflector from 1966 till 1991, and 370 of the R~CrB
measurements listed in Table~A2 were made with it.  It focussed the stellar
spectrum upon a physical diaphragm or mask, that looked like a high-contrast
photographic negative of a typical late-type spectrum, although it was
actually made artificially and was based upon the spectrum of Arcturus
(K2\hs2III) as portrayed in the {\it Arcturus Atlas\/} (Griffin 1968).  The
mask was scanned quite slowly in velocity space (about 1.4 \kmss per second)
along the direction of dispersion, and at a certain position, determined by
the stellar radial velocity, its transparent apertures were systematically
aligned with the absorption lines in the star spectrum.  At that point there
was a significant diminution in the total light that the mask transmitted,
which was gathered by a field lens (Fabry lens: Redman 1945) and measured by
a photomultiplier.  The output from the latter was drawn in real time by a
strip-chart recorder running at one inch per minute.  Passage through the
light-transmission minimum (`dip') took about half a minute, after which the
observer reversed the direction of scan by operating a switch, and could
thus make several passes backwards and forwards through the dip.  A typical
observation consisted of 6--10 passes.  During the scanning fiducial marks
were incorporated in the chart-recorder trace at certain velocities.  The
reduction procedure consisted of bisecting the dips by eye with a line drawn
on a small piece of perspex, noting their mean position with respect to the
fiducial marks and doing some arithmetic on a slide-rule.  On occasions the
dips given by R~CrB were somewhat asymmetrical but, since the bisection
procedure was subjective anyway, an asymmetry did not constitute any
particular problem, and the measurer was never conscious of there being any
serious ambiguities as to the positions of the effective centroids of the
dips\footnote{The dip profile can be regarded as the averaged profile, in
  the spectrum of the object observed, of the hundreds of absorption lines
  that are transmitted by the mask in the spectrometer -- mostly lines of
  neutral metals.  Of course it is not as informative as an actual spectrum
  -- a feature in the dip may be attributable only to a sub-set of the
  transmitted lines, for example just those arising from high excitation
  levels, and in that case the observed feature must in fact be
  correspondingly more marked in that set of lines from which it
  originates.}.

\begin{figure}

\begin{center}
 \includegraphics[width=3.5in]{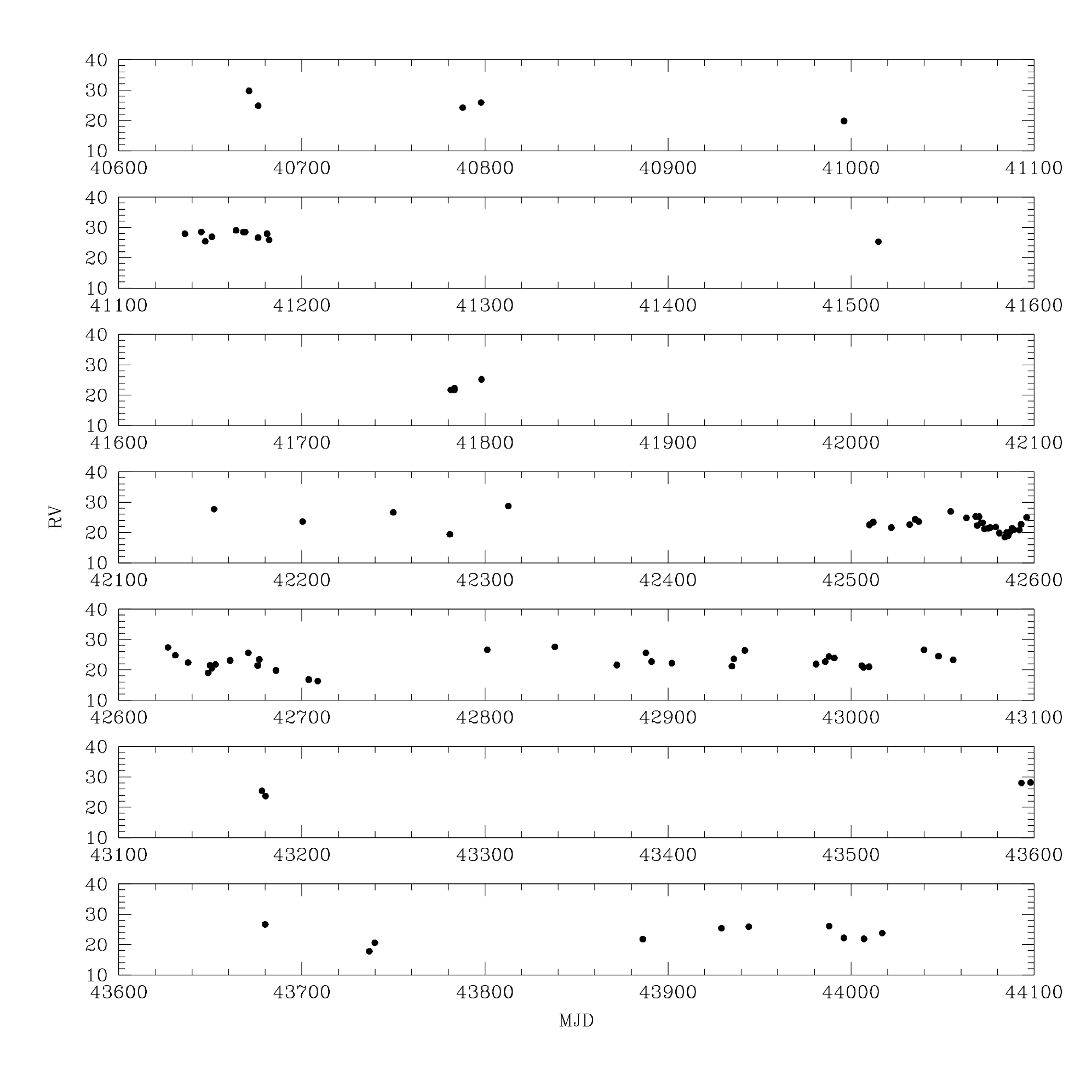} 
 \caption{The early Cambridge radial velocities (RV) as a function of MJD.}
   \label{rv}
\end{center}
\end{figure}

\setcounter{figure}{1}
\begin{figure}
%\vspace*{-4.0 cm}
\begin{center}
 \includegraphics[width=3.5in]{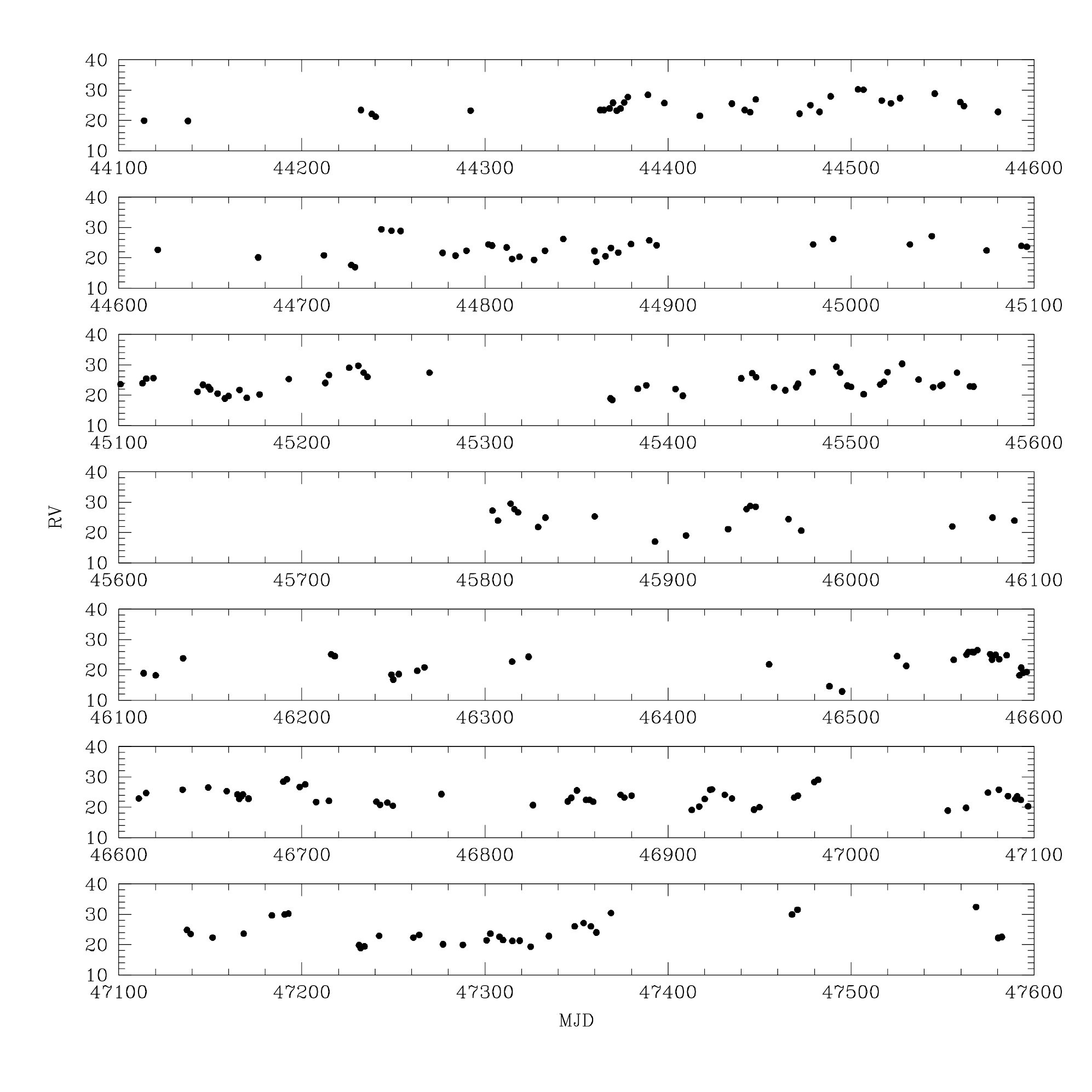} 
 \caption{continued. In section 3.4.2, point 3,  the period from MJD 45140 to 45160 is referred to as `section a' while that between 45214 and 45300  is called `section b'. }
\end{center}
\end{figure}

\setcounter{figure}{1}
\begin{figure}

\begin{center}
 \includegraphics[width=3.5in]{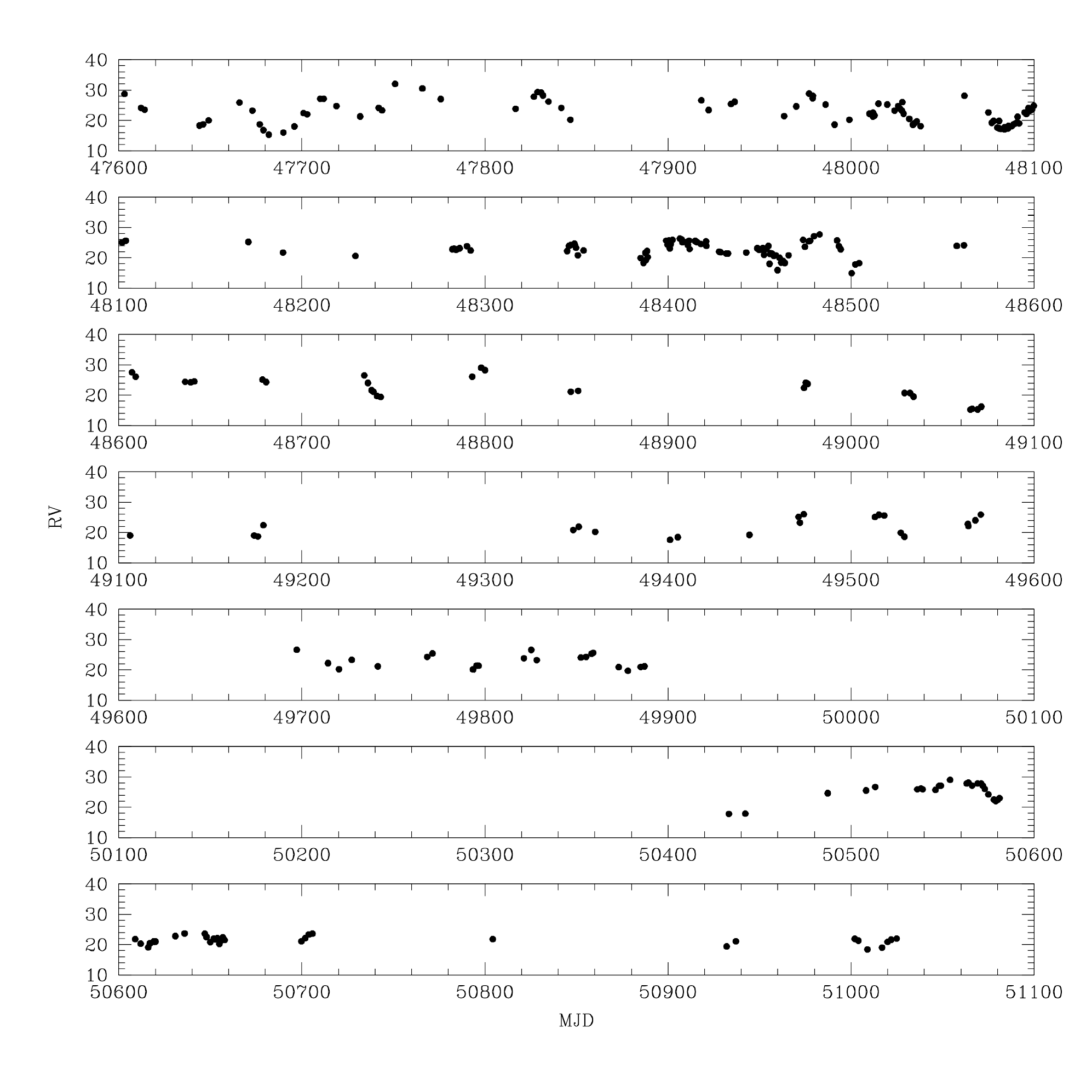} 
 \caption{continued.}
\end{center}
\end{figure}

\setcounter{figure}{1}
\begin{figure}
\vspace*{-7.0 cm}
\begin{center}
 \includegraphics[width=3.5in]{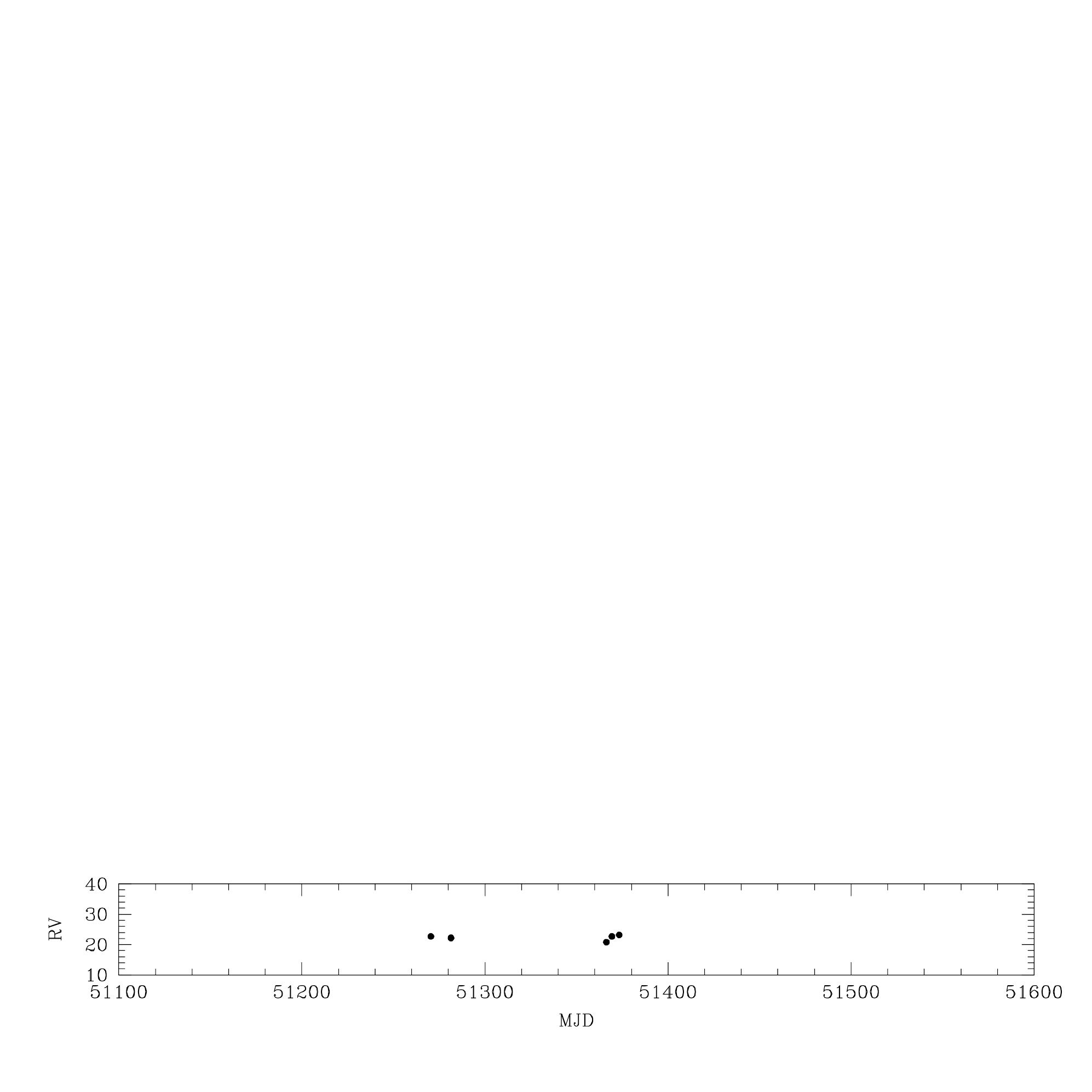} 
 \caption{continued.}
\end{center}
\end{figure}

Subsequent radial-velocity spectrometers that operated with physical masks
were computer-controlled, the first being the one that was constructed by
Griffin \& Gunn (1974) to operate at the coud\'e focus of the Palomar
Observatory's 200-inch telescope; it contributed just 4 measurements of
R~CrB to Table~A2.  In such instruments the scanning is performed relatively
quickly -- the whole scan is repeated at a frequency of a few Hz rather than
taking half a minute or more as in the prototype instrument.  The photon
counts are accumulated in a computer memory in a succession of `bins' at
addresses that are advanced in synchronism with the scan, so there is only
one picture of the dip but it builds up progressively as more and more of
the quick scans are overlaid; the observation is terminated by the observer
at will, the final array of bin counts is stored, and the bisection of the
dip is performed automatically by the controlling computer and/or
subsequently by more-sophisticated software.

RFG was fortunate to have many opportunities to use the Geneva Observatory's
\cors spectrometer (Baranne et al.~1979) on the 1-m telescope at
Haute-Provence (OHP), and a few on the similar instrument on the Danish 1.54-m
telescope at the European Southern Observatory (ESO); those instruments contributed 108 and 4 measurements,
respectively, of R~CrB.  He also made 15 observations of the star as a guest
observer with the spectrometer at the Dominion Astrophysical Observatory's
1.22-m coud\'e (DAO) (Fletcher et al.~1982).  All those measurements have been
included, suitably flagged, in Table~A2.  The largest contribution (541) to
the set of radial-velocity measurements discussed here, however, has come
from the \cors installed at the Cambridge 36-inch coud\'e focus in
succession to the original spectrometer.  It operated for a time in 1997,
and then from the end of 1999 until (as far as R~CrB is concerned) 2007,
when the star fell out of sight in a unprecedentedly deep and protracted
photometric minimum.  The 28 measurements made in the early months of 1997 have been included in Table~A2, while the main bulk of the Cambridge \cors data (513 nights) appear in
Table~A3.

In all the spectrometers operated by computers, the centroid of the observed
dip is determined by cross-correlation with a standard profile.  In the
Geneva Observatory's instruments that profile is a Gaussian, but in the
Palomar and Cambridge ones it starts as the profile that is given by the
instrument concerned for the many stars that prove to give dips of the
minimum width.  It is matched to the wider dips given by other stars by
broadening into a rotational profile by summing the contributions of a lot
of elementary areas into which the stellar disc is conceptually divided, to
each of which is assigned the velocity corresponding to its distance from an
axis about which rotation is deemed to occur and a brightness found from
the conventional limb-darkening model.  The observed dip is modelled in
position, depth and rotational velocity by adjusting those parameters
computationally to minimize the sum of the squares of the discrepancies of
the whole set of `bin' counts that represents the dip from the corresponding
elements of the model.  Of course the method is really designed to work with
symmetrical dips, since the model ones are necessarily symmetrical; but it
gives a sensible-looking result, much the same as a human measurer with his
line ruled on a piece of perspex would obtain, when it is called upon to
adjudicate on a skew dip.  Except as noted in specific cases below, the dips
given by R~CrB are not normally skew enough for there to be ambiguity at a
level at all comparable with the typical range of variation of the star's
radial velocity.

Table A3 presents not only the measured radial velocities but also, for each
observation, parameters related to the strength and width of the dip.  The
strength is given as an equivalent width, defined exactly as for
line-strengths in stellar spectra except (since the abscissae of a plotted
dip are velocities rather than wavelengths) the unit is the \kmss rather
than the \AA.  The dip width is also expressed as a velocity, in terms of
the \vsinis value of the best-matching rotationally broadened model dip.
In~R~CrB that value is normally between 15 and 25~\kms.  Although it is
useful to have a numerical quantification of the dip width, there is no
suggestion here that rotation is the principal cause of the dip width in
R~CrB.  We visualise the star as presenting to view a hemisphere containing
a rather small number of individually huge convective elements, and the dip
width is compounded of the turbulent velocities within each individual
element and the considerable differences among the several elements in
their mean motions in the line of sight.  Because we assess dip widths
by means of a rotational model, for convenience we refer to \vsinis values
below, but only as a shorthand way of quantifying the dip width and without
implying that the numbers truly refer to rotation of the star. Data for the times covered
 by Tables A2 and A3 are plotted in Figs~\ref{rv} and \ref{coravel} and are discussed in detail later.

\begin{figure}

\begin{center}
 \includegraphics[width=3.5in]{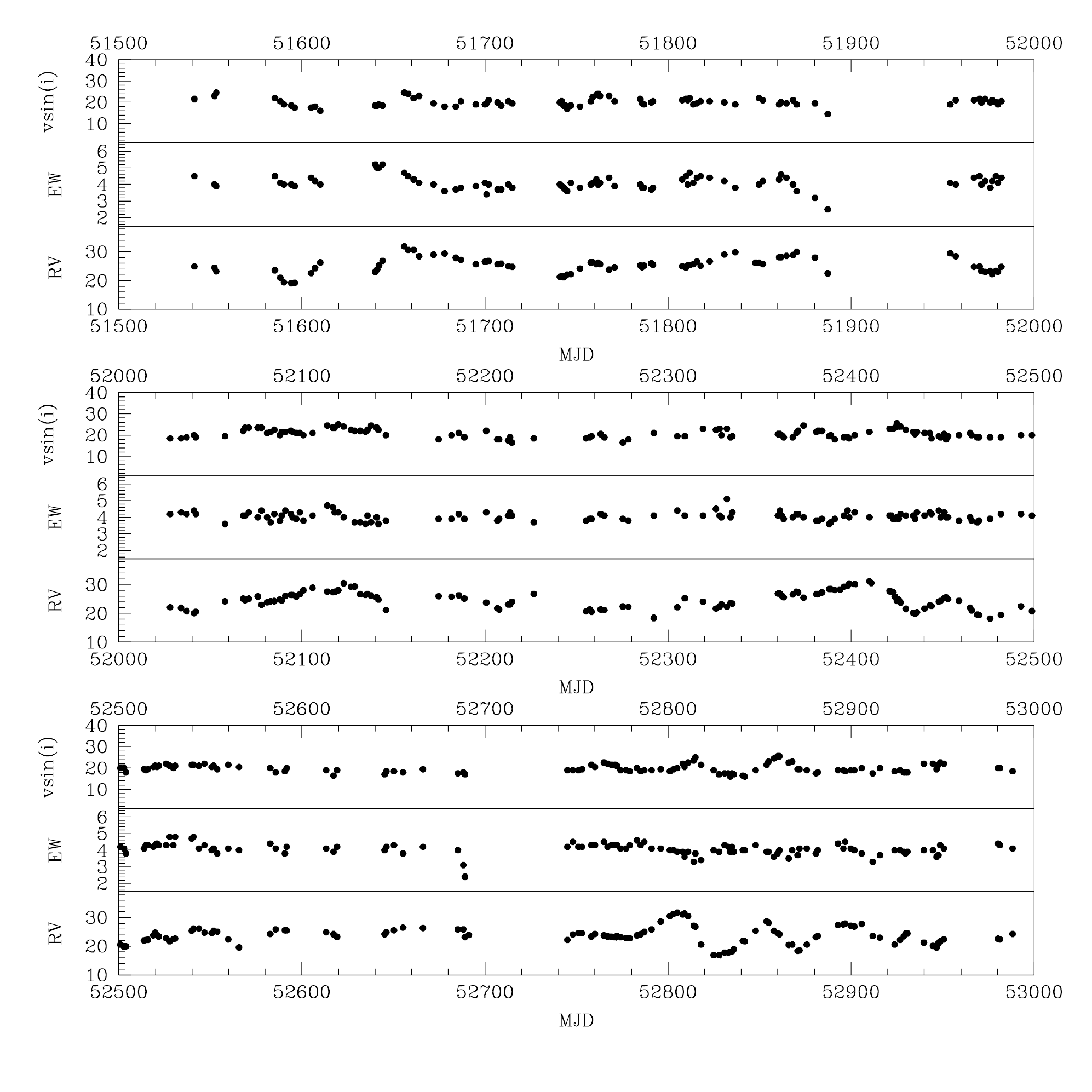} 
\caption{\cors data; the radial velocity (RV), equivalent width (EW) and $v \sin{i}$ are shown as a function of MJD. }
\end{center}
\end{figure}

\setcounter{figure}{2}
\begin{figure}
\vspace*{-0.5cm}
\begin{center}
 \includegraphics[width=3.5in]{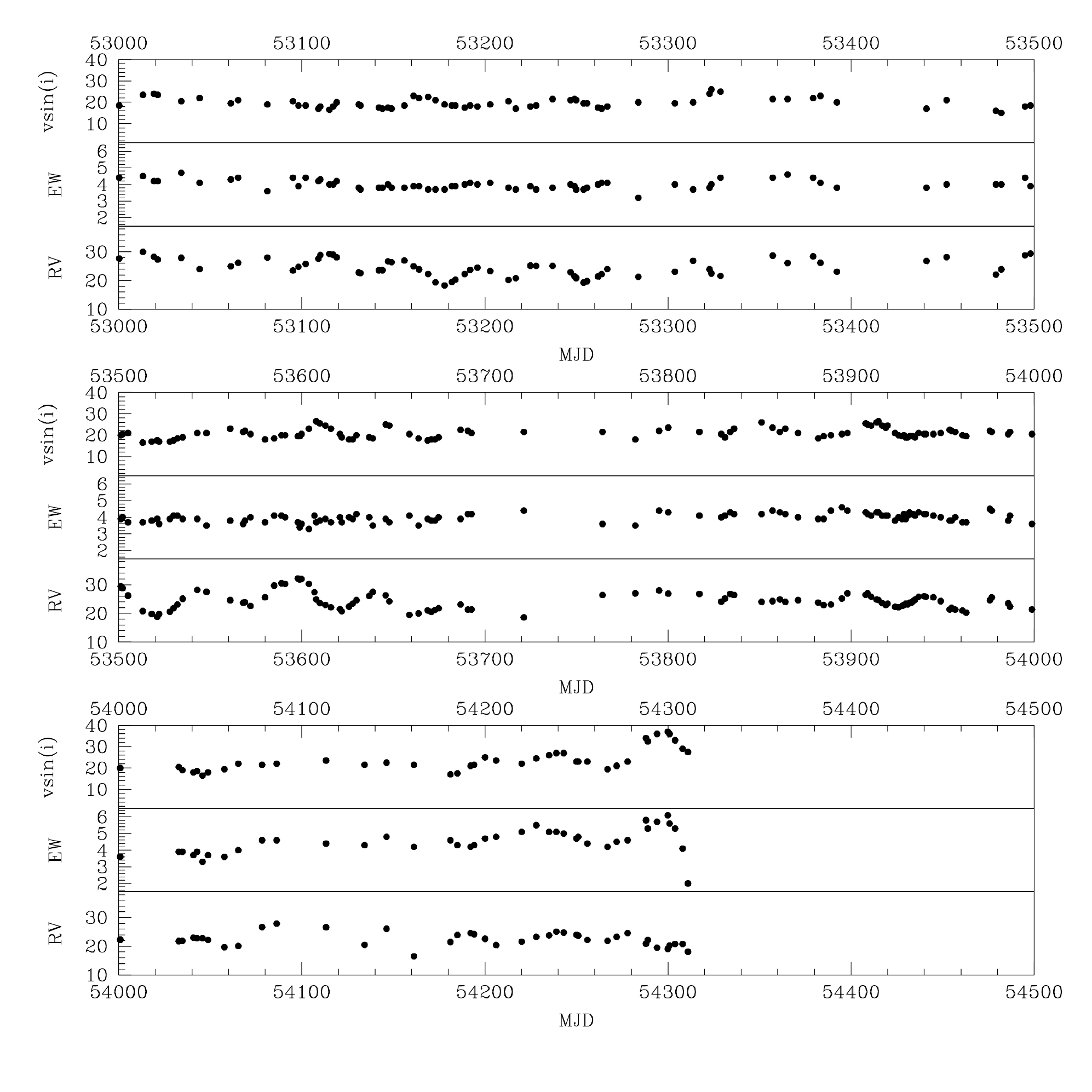} 

 \caption{continued. }

 \label{coravel}
\end{center}
\end{figure}

\section{Variations at Maximum Light}

\subsection{Previous Observations}

Before discussing our radial-velocity observations near maximum light, it is
useful to summarize previous work on the small-scale variations in both
light and radial velocity.

Since the discovery of its variability by Pigott (1797) there have been extensive visual
observations of R~CrB.  
Many of them have been assembled by the 
American Association of Variable Star Observers
(AAVSO).  They
are of importance in studying the major declines but are rarely of
sufficient accuracy to delineate the small-scale variations clearly.\footnote{The more recent AAVSO data contain a limited number of photoelectric and CCD
observations by their members.}
Several authors have obtained multi-colour photometry, and those giving
information on the small-scale variations are listed below together with a
brief summary of their conclusions.  Analogous comments are made on previous
radial-velocity observations.

Fernie et al. (1972).  {\it UBVR} observations in 1971 
(27 observations over 113 days) yield a period of
$45.4 \pm 2.2$ days.

Tempesti \& De Santis (1975).  Their data over 175 days can be fitted by a combination of
$44.41 \pm 0.44$ days and $26.80 \pm0.28$ days (see Fernie 1989).

Fernie (1982).  {\it UBVRI} photometry in 1971 and 1978--80 yield a
quasi-period of about 46 days.

Fernie et al. (1986).  {\it UBV} observations in 1985 indicate a period of
$45 \pm 5$ days just prior to an RCB event.

Fernie (1989). Five photometric $V$ maxima in 1985--87 fit a period of
$43.8 \pm 0.1$ days. 

Fernie (1990a). {\it UBV} on 99 nights in 1988 showed the usual $\sim$0.2-mag
amplitude in $V$ but no single period.  Periods of 59, 31 and possibly 23
days are present.

Fernie (1990b). {\it UBV} on 98 nights in 1989 as the star emerged from a
decline gave a period of $41.1 \pm 1.5$ days.

Fernie (1991). {\it UBV} observations on 71 nights in 1990 indicated
a period of $44.6 \pm 0.6$ days.

Lawson (1991) re-analysed photometry from the literature.  He found
that the 1986--89 data were best represented by a coherent periodicity
of 51.8 days, but in 1985 and 1990 there was a $\sim$44-day periodicity, of
which there was no evidence in 1986--89.

Gorynya et al. (1992) made 43 radial-velocity observations over 134 days in
1990 and 43 over 138 days in 1991.  Attempts at period-fitting to the entire
sample were unsuccessful.  The 1990 measurements could be fitted with a
period of 42.97 days and an amplitude of $\sim$10 \kms, but that did not
fit the 1991 observations, which gave a much less convincing periodicity of
33.97 days.

Fernie \& Lawson (1993) discussed {\it UBV} photometry and radial velocities in
1990 and 1991.  The 1990 data were possibly consistent with a 43-day period,
which was not seen in the 1991 photometry but might be present in the radial
velocities.
 
Fernie \& Seager (1994).  {\it UBV} observations of three maxima in 1992 and
three in 1993, taken together, fitted a period of $35.3 \pm 0.2$ days within
$\pm$3 days over a time interval of 12 cycles.  An RCB event in 1993 seems
to start near a minimum.  But the decline in 1985 (Fernie et al.~1986) seemed to
start near a maximum\footnote{Here, and throughout, maximum or minimum refer
to the cyclical variations near RCB maximum either in light or radial
\hbox{velocity}.}.

Rao \& Lambert (1997) deduced from spectra that the star was 500K cooler at
minimum than at maximum in a quasi-period cycle.  They also found that the
radial velocity, at least on one occasion near maximum light, depended upon
the excitation potential of the lines measured. Such behaviour had long
since been noticed, e.g.~Grenfell \& Wallerstein (1969), in Cepheid
variables, in which a shock wave periodically rises through the atmosphere,
affecting the spectral lines formed at progressively decreasing depths.

Yudin et al. (2002). {\it UBV} observations from 1985 to 1990 showed no stable
periodicity.

\subsection{Summary of photometric behaviour}

 Before turning more specifically to a discussion of the radial-velocity
behaviour of R~CrB we attempt to summarize the evidence about its
photometric behaviour.  As reviewed in part above, there is a large amount
of published photoelectric photometry in $V$ and other bands.  It tends to
be limited to a few months at a time and generally shows variations of about
0.2 mag in $V$.  Frequently the light-curves show cycles with intervals of
40 days or so between maxima, though sometimes longer or shorter cycles
are present.  The light-curve shapes do not repeat. Most of the observations are limited to an observing season of about 
100 days.

The
general conclusion from the observations is that both light and radial
velocity tend to show low-amplitude variations on time scales of the order
of 40 days in any given season, but the `period' is quite variable.  In view
of the different cycle lengths it is not surprising that no evidence has
been given that the variations keep phase from one season to another over
long time scales, or that any coherent periodicity can be traced over a
significant length of time. It is of course, not possible
to rule out the hypothesis that the star is undergoing regular coherent pulsations
at a level too small to be detected in the available data. 

\begin{figure}

\begin{center}
 \includegraphics[width=3in]{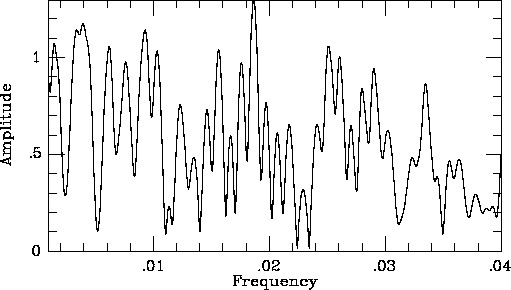} 

 \caption{Periodogram of the Lick velocities (omitting the 1956 point). }
   \label{ft_herb}
\end{center}
\end{figure}

\begin{figure}

\begin{center}
 \includegraphics[width=3.0in]{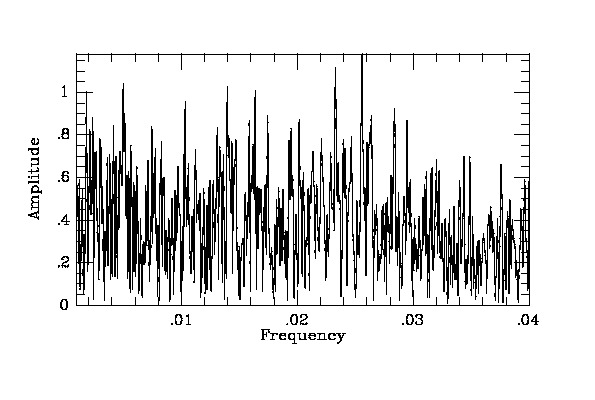} 

 \caption{Periodogram of all velocities in the time period MJD 39970 to 51374 (omitting the faint phases).}
   \label{ft1}
\end{center}
\end{figure}
\begin{figure}

\begin{center}
 \includegraphics[width=3.0in]{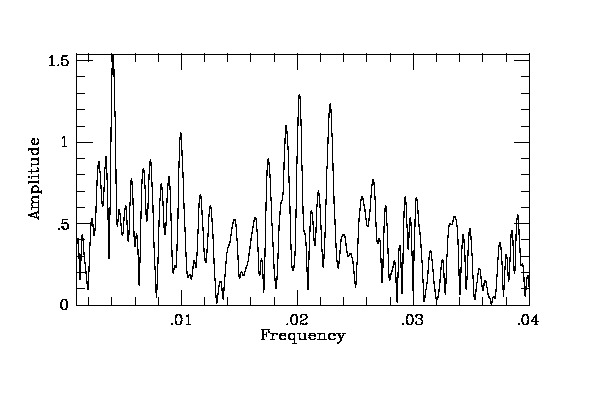} 

\caption{Periodogram of the \cors velocities (omitting the faint phases). }
   \label{ft2}
\end{center}
\end{figure}

\subsection{Search for periodicity in the radial velocities}
\subsubsection{Periodograms}

Figures \ref{ft_herb} to \ref{ft2} are periodograms of the data over different time intervals, Fig.~\ref{ft_herb} is for the Herbig data (except for the isolated 1956 observation).  
Fig.~\ref{ft1}  is for the 1968-1992 Cambridge data together with other
data from the literature around that time. 
Observations during RCB dips have been omitted.
 
Note that where
there is overlap in time between the Cambridge observations and others (e.g. the
Gorynya et al. (1992) observations), there is no evidence of zero-point differences. 
Fig.~\ref{ft2} is for the Coravel velocities, again with data during RCB dips
omitted.  In the last two cases plots including the RCB-dip phases are not significantly different. There are no outstanding
peaks in any of these plots which would indicate a significant coherent periodicity.
All the peaks are small and in any case differ from series to series,
e.g. 53.7 days in the Herbig plot,  39.0 and
43.0 days in Fig.~\ref{ft1}  (which includes the Gorynya et al. data), and
49.4, 44.2 and 52.6 days in the Coravel series.
Some evidence for a low-frequency component ($\sim 250$ days) in the Coravel series can be traced in 
the plots of the data against time, but its significance is not clear. 
   
\subsubsection{Wavelet analysis}
 The variability of R~CrB is obviously not strictly periodic and we therefore decided to apply a wavelet analysis 
to look for coherent behaviour that is only transient. To do this we used the WinWWZ package\footnote{
WinWWZ was produced by G. Klingenberg and L. Henkel; it can be downloaded from http://www.aavso.org/winwwz.} available through the AAVSO. This makes use of 
the weighted wavelet Z-transform (WWZ) as described  by Foster (1996). 
The wavelet employed is a simple sinusoid plus constant term, applied by sliding a window of predetermined width across the data. Points close to the centre of the window have the highest weight, whereas those near the edges have less weight.
The results are output as 3D plots of frequency/time/power.
This is particularly useful to reveal periodicities that are only present for a single season.

Note that the temporal coverage is uneven, times of intense ($\sim$ nightly) observations being interspersed  by more sparse observing. 
Some epochs of sparse coverage are inadequate to search for short periods ($<70$ days), and aliasing is potentially a problem.

\begin{figure}
\vspace*{-0.5 cm}
\begin{center}
 \includegraphics[width=3.5in]{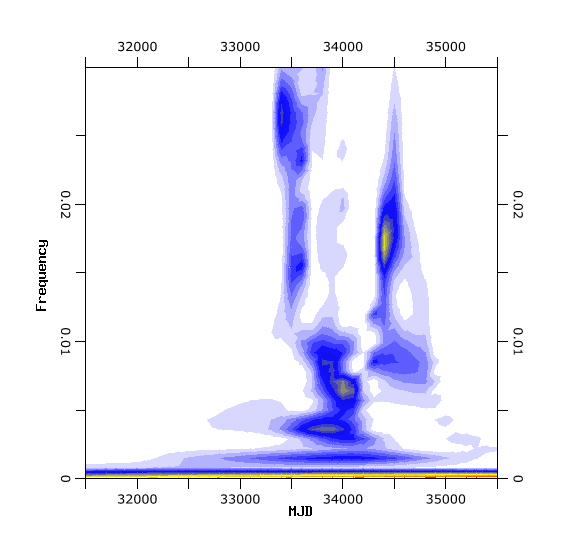} 

 \caption{(a) Wavelet analysis in which frequency (cycles $\rm day^{-1}$) is plotted against MJD. The density of the plot indicates where significant power is present at the given date and frequency.}
   \label{wave1}
\end{center}
\end{figure}
\setcounter{figure}{6}
\begin{figure}
\vspace*{-0.5 cm}
\begin{center}
 \includegraphics[width=3.5in]{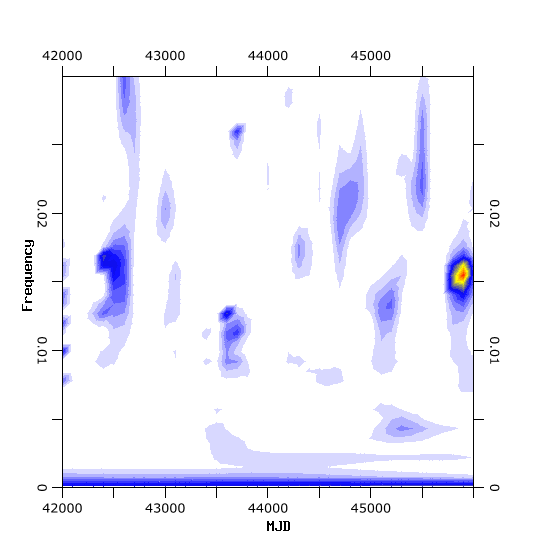} 
% \vspace*{-1.0 cm}
 \caption{(b) Continued.}
  %\label{wavelet}
\end{center}
\end{figure}
\setcounter{figure}{6}
\begin{figure}
\vspace*{-0.5cm}
\begin{center}
 \includegraphics[width=3.5in]{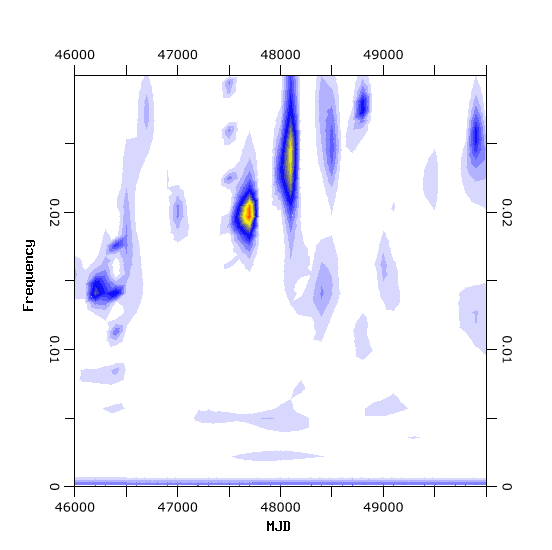} 
% \vspace*{-1.0 cm}
 \caption{(c) Continued}
  %\label{wavelet}
\end{center}
\end{figure}
\setcounter{figure}{6}
\begin{figure}
\vspace*{0.5 cm}
\begin{center}
 \includegraphics[width=3.5in]{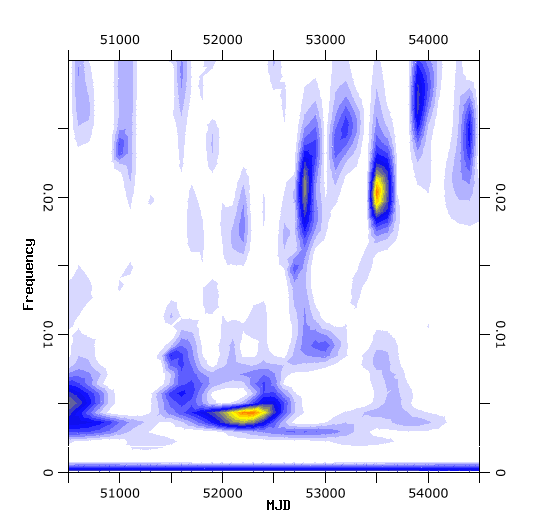} 
% \vspace*{-1.0 cm}
 \caption{(d) Continued.}
  %\label{wavelet}
\end{center}
\end{figure}

 Fig. \ref{wave1}a,b,c,d 
show the WWZ transform over four time periods. The diagram is
blank when there are no data or there is no periodicity in the data.

 It is clear that no single period persists through several seasons. In the early data a period of 50 to 60 days (0.016 to 0.019
$\rm day^{-1}$) is present around MJD 34400 %(Fig. 7a) 
(Fig.~\ref{wave1}a).
At MJD 47700 and 48100 periods of 50 days (0.02 $\rm day^{-1}$) and 40-45 days (0.024 $\rm day^{-1}$), respectively,  are prominent. 
There is an indication that the period changes gradually over four seasons
from about 50 days at MJD47700 to about 36 days at 48800 i.e. from 0.02 to 0.028 $ \rm day^{-1 }$

At MJD 50400 and 52400 (Fig.~\ref{wave1}c) periods of 200 and 232 days dominate, but may be artifacts of the gaps in the data.

After 52700 the dominant cycle time seems to take various values between 36 and 50 days (0.028 and 0.02 $\rm day^{-1}$) over the last four seasons, as illustrated  in  Fig.~\ref{wave1}d.

\subsubsection{A re-examination of earlier data}
\begin{figure}
%\vspace*{-4.0 cm}
\begin{center}
\includegraphics[width=3.5in]{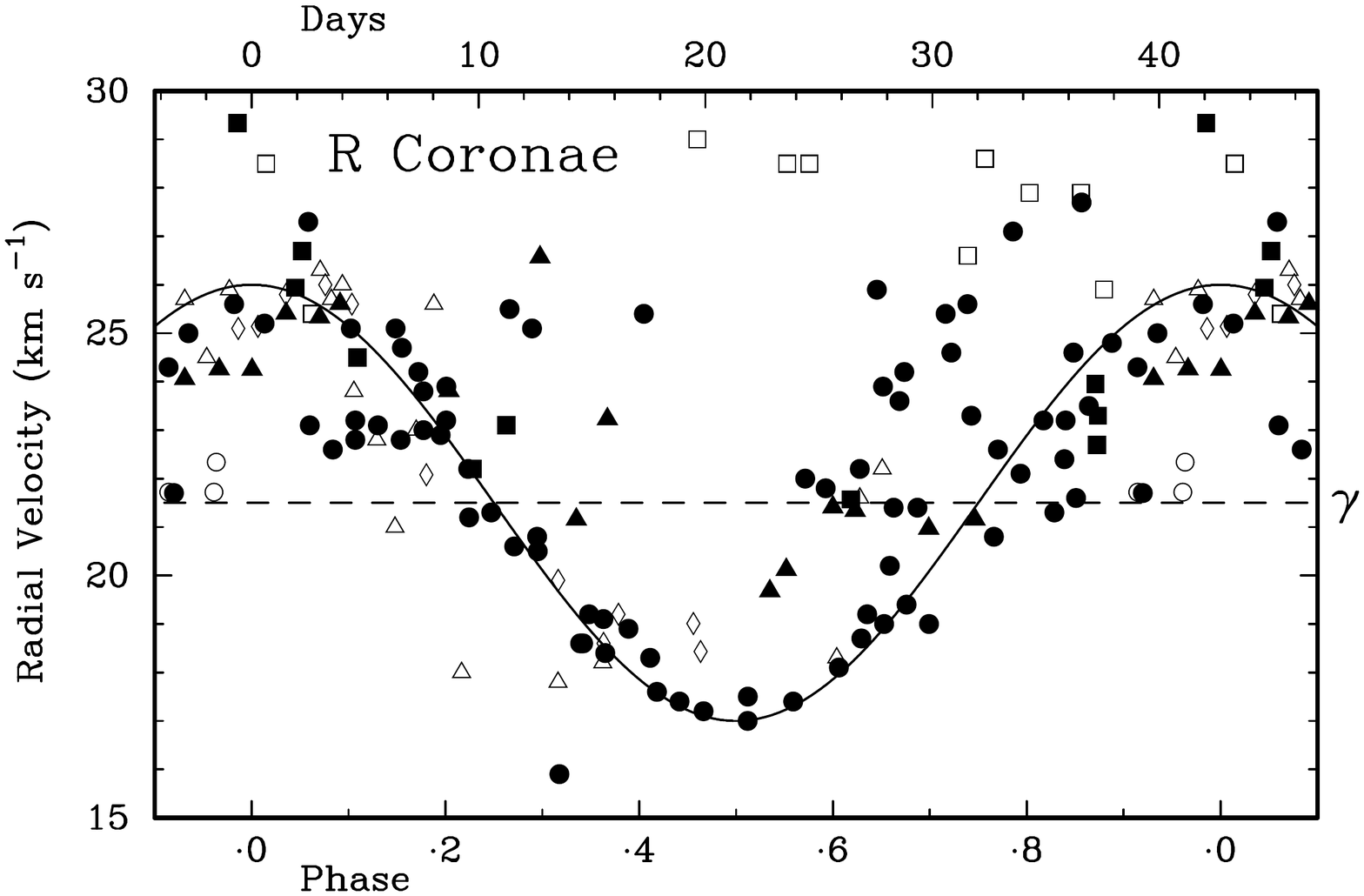} 
% \vspace*{-1.0 cm}
 \caption{Data discussed by Rao et al. phased on a 42.7 day period; the solid circles are the Gorynya observations.}
  \label{griffin}
\end{center}
\end{figure}

Rao et al. (1999) carried out a periodogram analysis of their own
radial velocities and ones collected from the literature and 
suggested a period of 42.7 days. Professor Rao has very kindly sent us a
list of the velocities used in their analysis.  Besides his own, they are
primarily from the papers mentioned in section 3.1 (Gorynya et al.~1992,
Fernie \& Lawson 1993, Rao \& Lambert 1997), with a small number of
observations from Keenan \& Greenstein (1963), Fernie et al.~(1972) and Rao
(1974).  A periodogram of these data  
shows a peak near 42.7 days but this is by no means outstanding.  Most
of the series used in the analysis consist of rather few observations; the
data are in fact dominated by the observations in 1990/91, particularly the
Gorynya et al.~series which is among those noted above.  Fig.~\ref{griffin} shows the
data used by Rao et al.~plotted on a 42.7-day period and with the various
observations distinguished.  It is evident that the Gorynya data dominate the
diagram and  figs 1 and 2 of their paper show, as already mentioned, that while there is clear cyclic variations
with a ``period" of 42.97 days in 1990, this is not present in 1991 which gives an uncertain ``period" of 33.97.
%EVIDENT THAT THE GORYYNA DATA DOMINATE
%THE DIAGRAM  AND FIG1 AND FIG2 OF THEIR PAPER SHOWS THAT WHILE THE 1990 SHOW A CLEAR CYCLICAL VARIATION (42.97 DAYS)
%THEIR 1991 VELOCITIES ARE VERY SCATTERED AND GIVE AN UNCERTAIN "PERIOD" OF 33.97DAYS.
It is clear that the observations in 1990  are sufficiently numerous to commandeer the
periodogram.  There is little if any sign of such a period in the other
data.  Evidently whilst cycle lengths of about 40 days have been seen from
time to time (as summarized in section 3.1), the data analysed by Rao et
al.~do not demonstrate that that is a coherent periodicity.

Crause et al. (2007)  have suggested that the RCB-type declines of R~CrB
occur as multiples of a very precise period ($42.97 \pm 0.03$ days) which they
take as a pulsation period. The above results and those discussed in earlier sections
show no significant  coherent, long-term, signal at that period. 
The question is further discussed in section 6  in connection with the nature of
RCB-type declines.
\subsubsection{Conclusion}
The conclusion from all the sets of data is that there appears to be no true
periodicity in the radial-velocity measurements, any more than there is in
the photometric ones, although R~CrB does often show light and velocity
variations having a characteristic time-scale of the order of 40 days.

\subsection{Further discussion of radial-velocity variability}

Although, as just discussed, the radial-velocity data give no indication of a
sustained coherent periodicity, it will be evident from Figs~\ref{herbig} to \ref{coravel} that at some
epochs there is evidence for cyclical variations,
as is also clear from the wavelet analysis (Figs~\ref{wave1}).  The following are further brief
notes on those variations as seen in the three series of data. They are in
chronological order in each series.\\

\subsubsection{The 1950--1953 series}
  %This is mainly the Herbig data with a few observations from
%Keenan and Greenstein (1963).\\ 

 A plot (Fig.~\ref{herbig}) of the Herbig series shows a
velocity range of 19 to 28 $\rm km\,s^{-1}$. 
%The straight mean of all the velocities is --.  
The star was in a major decline from about MJD
34300 to about 34450.  The figure gives the impression that the
velocity was slightly lower in the first set of observations (roughly
33322--33550) than in the next set (roughly 33650--33900), i.e. before and after
the decline.
 
Clearly that could be affected by the distribution of the points.

 The WWZ transform  (Fig.~\ref{wave1}a) indicates periodicity around 50 to 60 days (about 0.017 $\rm day^{-1}$) in the data after the decline (MJD\,34400),  but nothing else really  stands out during this time (MJD\,33300 to 34600).\\ 

\subsubsection{The 1968-1998 series}
  This is primarily the older Griffin data with other observations, including those of
Gorynya et al. 1992, mentioned in section 3.3.3.

1.  MJD 42500--42710 (see Fig.~\ref{rv}).  The star was fading at the time of the last two
points. The observer estimated visual magnitudes of 7.5 and 8 on days 42703
and 42708 and the velocity was evidently declining at that time.  At the time
of the next observation, day 42801, the star was estimated at 8.6 mag and on
the rise.  Before the decline, the star shows a cyclical variation between
18 and 27 \kms.  The variation is obviously not regular, but a cycle length
of about 60 days is indicated.  It would be difficult to fit the
observations with a cycle length near 40 days.

2.  MJD 44800--44900.  As in the last paragraph there is evident irregular
cyclical variation (19 to 26 \kms ).  The time scale here is roughly 40 to
45 days.

3.  MJD 45140--45240. From 45140 to 45160 (section a) the star was near
normal brightness, but between about 45214 and 45300 (section b)  it went through
a very shallow RCB-type decline, being faintest (6.8 mag) on days 45226--45231 as
estimated by RFG.  
The velocities in section~a show little sign of regular
variability.  In section~b some rudimentary regularity is seen.  A cycle
with a period of about 40 days might be present.  The most noticeable
effect, however, is that the star's mean velocity in section~b is higher
than in section~a.  Alternatively, if section~a covered a velocity minimum
and such was missed in section~b, then the velocity amplitude is unusually
high ($\sim$10 \kms ) in the latter section.

4. MJD 45368-45567. These observations which  ended at the start of a decline
show two clear velocity minima separated by $\sim 43$ days
and the WWZ transform shows a weak peak from 0.021 to $>0.025\, \rm day^{-1}$,
which must be the same feature  
 (Fig.~\ref{wave1}b).

5. The WWZ transform shows a periodicity around MJD45900 of about 65 days
(0.016 $\rm day^{-1}$ ) 
(Fig.~\ref{wave1}b).

6.  MJD 46400--46800.  At the start of this period the star was undergoing a
shallow RCB event (days 46456/46489/46496: 6.7/7.5/7.8 mag (RFG)) and the velocity
is unusually low.  According to the AAVSO light-curve the star was back
close to maximum by MJD 46560. Between MJD 46540 and 46720 there is some
evidence of cycles but with no definite period.

7.  MJD 47200--47900.  An RCB event began about MJD 47360.  The velocities
on the initial drop (days 47361/47369: 6.6/7.5 mag (RFG)) are not low (compare
item 6 above).  By MJD 47568 the star was on the rise (47569/47603: 8.5/7.3 mag (RFG)).
R~CrB came out of a deep decline and was restored to normal brightness
$\sim$47620, but only about 100 days later (MJD 47743) began a decline to a short
moderate dip, reaching $\sim$8.5 mag $\sim$47775.  The velocities during
that period are covered in Fig.~\ref{rv}.  The star is likely to have been
slightly below normal light between the two dips. 

 The WWZ transform shows a period of around 50 days (0.02 $\rm day^{-1}$) between these
two fading episodes, centred around MJD 47700 %(Fig. 7c).\\ 
(Fig.~\ref{wave1}c)
During the earlier
part of this period Fernie (1990b) has {\it UBV} observations.  His data
cover two successive maxima and he derived a `period' of $41.1 \pm 1.5$
days, ``not significantly different from the customary 43.8 days''. 
The radial velocity data shown in Fig.~\ref{rv} might be fitted to a period near 41 days
between MJD 47640 and MJD 47730, though with varying mean velocity. 
However, the earlier and later points (just after and just before
declines) would not fit this period and the whole range is consistent with the
WWZ result.

8. MJD 47900--48500. This region contains other observations besides the Cambridge one,
in particular those of Gorynya et al. The  period of 42.97 days (Gorynya et al.) is based on the 
observations from MJD 47970 to 48104.
The WWZ transform shows a strong peak around 48100 extending from 0.022 to 0.026
$ \rm day^{-1}$. 

\begin{figure}
%\vspace*{-4.0 cm}
\begin{center}
 \includegraphics[width=3.3in]{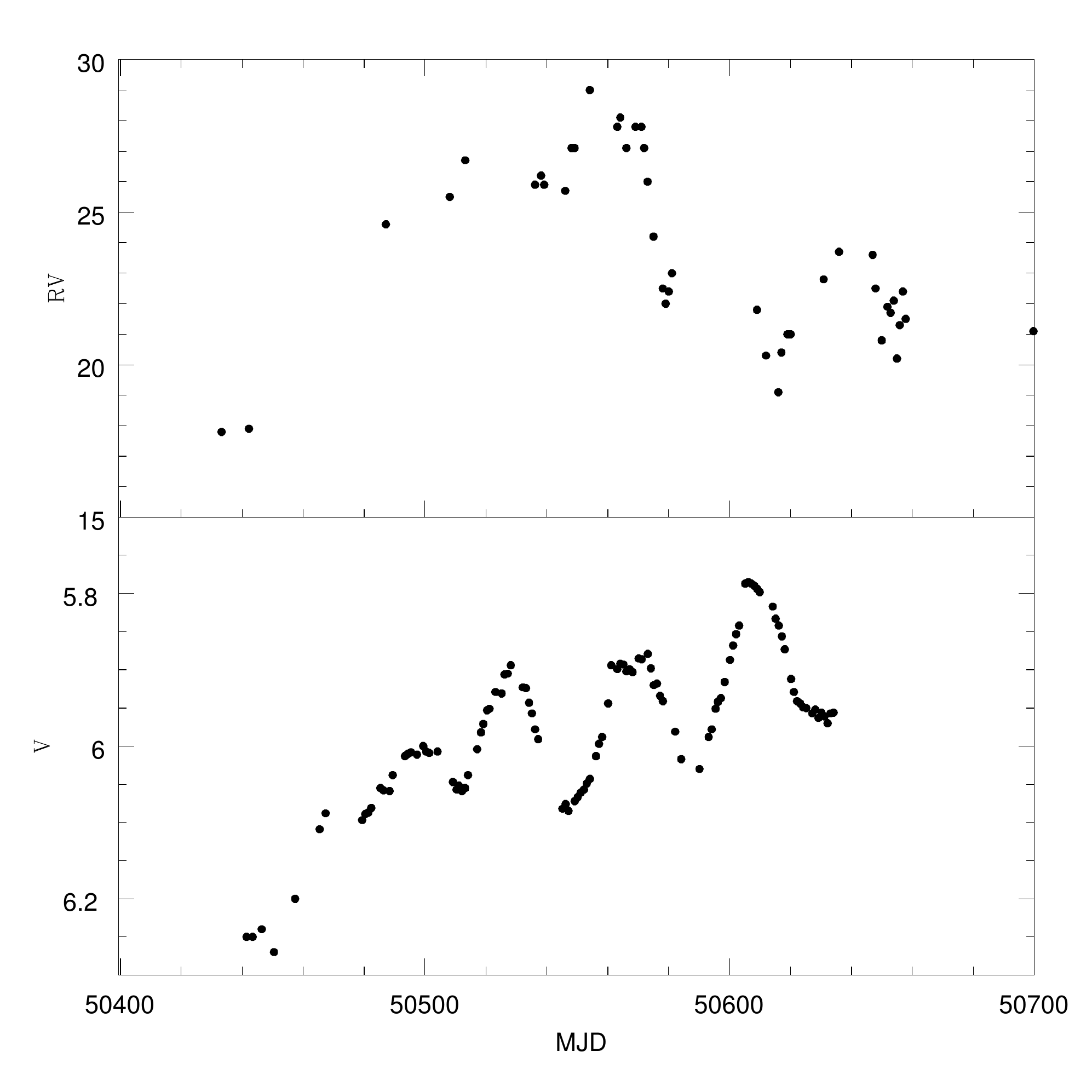} 
% \vspace*{-1.0 cm}
 \caption{$V$-band photometry (Yudin et al. 2002) compared to quasi-simultaneous radial velocities.}
   \label{yudin}
\end{center}
\end{figure}

9.  MJD 50400--50700.  The data are shown in Fig.~\ref{yudin}.  The $V$ observations
are from Yudin et al.~(2002).  At the start of this interval R~CrB was
slowly completing its recovery from a deep decline, returning to normal
brightness about day 50500.  The photometric data show cyclical behaviour, with
three maxima and three minima and a `period' of roughly 40 days
superposed on a general rise.  The
radial velocities can be regarded as showing a cyclical variation of a
similar length with, however, the mean velocity decreasing with time.
The periodicity is seen in the WWZ transformation (Fig. 7d) which also shows power 
at a much longer ``period" due to the long-term variation.
As in the interval just discussed,
light maximum occurs after velocity maximum and about the time when the
atmospheric material begins moving away from the star.

\subsubsection{The 1999--2007 series} 
 All the observations discussed in this section were obtained with
the \cors instrument.
The \cors data yield measures of radial velocity, equivalent width (EW) and
a measure of line width that is expressed as \vsinis as explained at the end
of \S2.2 above. These are plotted in Fig.~\ref{coravel}.  In~this section we
briefly summarize some aspects of these plots which are of relevance for the
question of cyclical behaviour. Other aspects will be considered in later
sections.

1.  MJD 51500--51900 (Fig.~\ref{coravel}).  At the start of this period the star was on
the rise from an RCB-type decline.  From about 51590 to 51860 it was near
maximum, though about 0.2 mag below its usual maximum brightness.  It~began
dropping again at the end of this interval.  Between 51640 and 51710 the star
showed large variations in all three quantities, though with no clear
periodicity.  Radial velocity and \vsinis have their maximum values at
closely the same time.

 The large variations in EW just before a drop to minimum in 2007 will be
discussed below. The variations in the interval under discussion are somewhat
smaller, but still well above the average. In view of the 2007 discussion
this can be taken as an indication that the type of behaviour discussed in
\S4.2 is not necessarily followed closely by a decline.

At the end of
this interval EW falls as the star goes into a decline.  This effect is seen at
other declines and is probably due to the filling-in of the lines by the emission
spectrum which dominates when the star falls further in light.

2.  MJD 52400--52500.  During this interval, when the star was
well away from declines, there is an apparent cyclical behaviour in RV
(three maxima, two minima) of about 41-day length superposed on a marked
longer-term decrease in mean velocity.The periodicity is weakly seen in the
WWZ transformation (Fig. 7d) where most of the power is in the longer-term
variation. Any periodicity immediately before or after this interval
must, if present, be of much smaller amplitude.
The varying mean
velocity is hardly compatible with a simple radial-pulsation model.

3. MJD 52680--52700.  Typical decrease in EW as the star goes into a decline.

4.  MJD 52740--52960.  At the start of this period the star was brightening
from an RCB decline.  It was back near normal brightness at $\sim$52760.  Two
cycles of large RV range with a cycle length around 45 days are clearly
seen
 and show in the WWZ transform at 0.021 $\rm day^{-1}$.
At the same epoch, \vsinis shows similar variation with maxima
following RV maximum by a few days.

5.  MJD 53000--53500.  Variations in RV on a time scale of the order of 40 days
are clear during at least parts of this period.
They are shown in the WWZ transform (0.025 $\rm day^{-1}$) centred at 53200 %(Fig. 7d). 
(Fig.~\ref{wave1}d).

6.  MJD 53500--53700.  There is clear evidence of cyclical behaviour of RV
(3 cycles).  The interval between the two clearly defined minima (53521,
53621) is 50$\times$2 days
 and shows very strongly in the WWZ transform at 53500 with 0.02 $\rm day^{-1}$
as already noted above.  
   However, the shape and amplitude of the velocity
curve clearly changed from cycle to cycle.  As before, \vsinis has maxima
slightly later than RV.

Further discussion of these data is given in \S4.

\subsection{Summary of variability at maximum light}
The results discussed above may be summarized
as follows:

1.  There are times when cyclical behaviour on a time scale of $\sim$40 days
is evident in the radial velocities and sometimes also in $v \sin i$.  The
amplitudes and velocity-curve shapes can, however, differ markedly from
cycle to cycle.  Earlier work cited in \S3.1 is consistent with those
findings.

2.  Periodograms of long stretches of velocity data as well as a wavelet analysis show no dominant
periodicity.

These results seem to rule out a coherent-pulsation model for the
observed variations in R~CrB.  We discuss a physical model for these results below, after
first reviewing the \cors results in more detail.

\section{Line shapes in the coravel series}

The \cors series is particularly valuable because it gives
the quantities analogous to equivalent width (EW) and broadening
(\vsini )
of the profile (the `dip'). The variation of these quantities has been discussed 
above (\S3.4.3). 
In addition the profiles of the dip were recorded.
These are particularly relevant to our understanding of the RCB phenomenon and
in the following sub-sections we concentrate on variations in the line
profiles and their implications for phenomena in the stellar atmosphere and
the formation and ejection of carbonaceous particles. 
%\bf{Fig 9 is a montage of some of the data.} 
%THIS SHOULD BE PUBLISHED FULL PAGE SIZE SINCE THE PLOTS ARE SMALL.  THE CAPTION SHOULD SAY THAT IN THE
%TEXT THE COLUMNS ARE DENOTED A,B,C,D LEFT TO RIGHT AND THE ROWS 1 T0 10 TOP TO BOTTOM. IN A1 AND B1 THE YEAR SHOULD BE 2005 NOT 2003}
\begin{figure*}
%\vspace*{-4.5cm}
\begin{center}
\includegraphics[width = 6.5in]{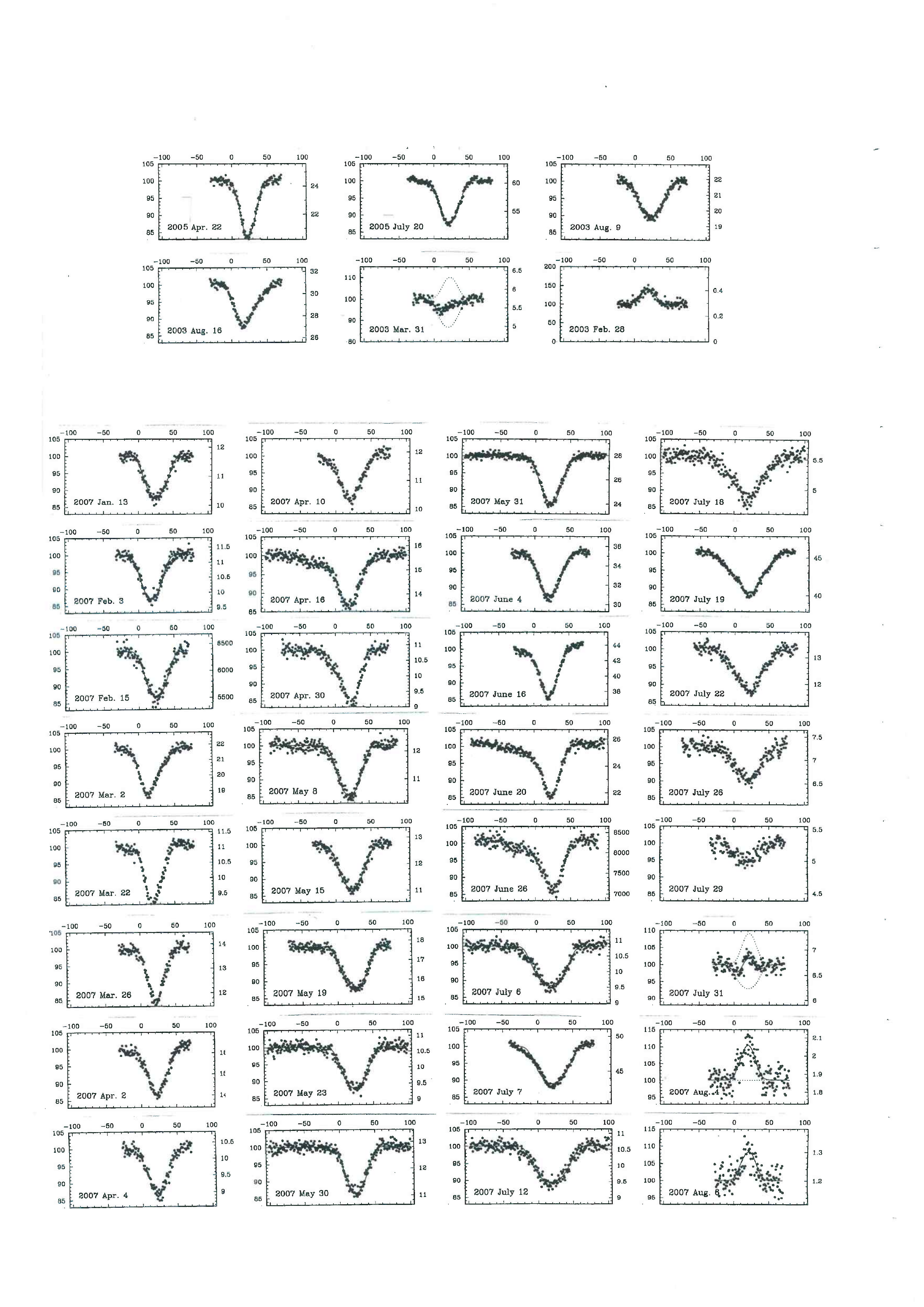} 
%\vspace*{-0.5 cm}
\caption{A montage of some of the \cors profiles. These are discussed in detail in section 4, where the rows are numbered 1 to 10 and the columns lettered A to D, so that, e.g., the profile for 20 June 2007 is referred to as Fig. 10 (6C). }
%Note that for profiles 1A and 1B the year should be 2005 not 2003}
\label{montage}
\end{center}
\end{figure*}
We first discuss examples of typical profiles before passing to a detailed account of the great decline of 2007
and its interpretation. We then discuss further examples of profile changes and their
interpretation.

\subsection{Typical profile behaviour}

 The next two sections summarize ``typical, normal" behaviour at maximum and in decline. In later sections
the line shapes at particular epochs  and their interpretation are discussed. 
At other times during the \cors
series the line shapes are within the range discussed in \S4.1.1
and are not noticeably asymmetric.

\subsubsection{Typical profiles at maximum}
Fig. 10, (1A), (1B) and (1C) illustrates the normal range of \cors profiles near maximum light.
1B is a typical profile whilst 1A and 1C
show the extremes of width normally encountered. 
These profiles are all symmetrical:\\
(a) Fig. 10 (1B) MJD 53572 (20/7/05) EW = 4.0  : \vsini = 20.5,\\
(b) Fig. 10 (1A) MJD 53483 (22/4/05) EW = 4.2 : \vsini =15,  \\
(c) Fig. 10 (1C) MJD 52861 (09/8/03) EW = 4.0 : \vsini =25.5 \\
Changes in the profile around maximum are discussed in section 4.4.

\subsubsection{Profiles during an RCB-type decline.}
(a) Fig. 10 (2C) (28/2/03) MJD 52700.  The star was $\sim$12 mag (RFG) and nearing the
faintest point of an RCB decline, having dropped from maximum by
$\sim$6~mags (visual) in $\sim$20 days.  The spectrum is obviously dominated
by emission lines (which have been frequently reported during early
declines of R~CrB and other stars of this class (e.g. Herbig 1949)).  The
emission spectrum is believed (e.g. Payne-Gaposchkin 1963) to be a permanent
feature arising in the outer atmosphere (sometimes referred to as the
chromosphere) and seen only when the main body is eclipsed.  Its variation
at an eclipse by an expanding puff of soot was discussed by
Alexander et al. (1972) and Feast (1979) in the case of the RCB star RY
Sgr.

(b) Fig. 10 (2B) MJD 52731 (31/3/03).  This profile was observed 31 days after the previous
one.  The star was then on the rise towards maximum with an estimated
magnitude of 9.5 ($\sim$3.5~mag below maximum). The line can be fitted, as
shown by the dotted lines, by the combination of emission and absorption
lines and with the absorption being displaced to the blue.  During the declining phases of
the 1948--49 and 1960 RCB events, Herbig (1949) and Payne-Gaposchkin 1963)
found evidence for a shift in the opposite sense.

\subsection{The Great Decline in 2007: MJD 54163--54319}
\subsubsection{Observations}
In 2007 July, $\sim$MJD 54291, R~CrB went into a rapid decline from which it did not fully recover until long after the interval considered here. 

In this section we report and
discuss the \cors data for the 130 days leading up to the decline and the
first 20 days of the decline itself.

A striking feature shown in Fig.~\ref{coravel} is the variation of \vsinis
and EW as well as the line shapes at that epoch.  Fortunately there is some,
limited, high-dispersion spectroscopy (Rao \& Lambert 2010, = RL10) which is
valuable in aiding the interpretation of the \cors data.
The \cors line profiles show that complex changes were
taking place on a time-scale of a few days. 

The following is a brief summary of changes taking place during this time together with results from RL10.\\
Fig. 10 (6A)  (2/3/07)  MJD 54161. The line profile is not symmetrical.  The profile would be
reproduced by a symmetrical line fitted to the lower part of the line\footnote {Here and
  throughout, blue/red refer to the low/high velocity sides of the profile.},
together with a shallow component on the red side the profile centred at
$\sim$40 \kmss ($\sim$25 \kmss from the mean). \\
MJD 54163. RL10 write ``Star appears disturbed with emission in some lines
suggesting a component with an inverse P Cygni profile.''
This description, particularly the suggestion of additional red-wing
absorption, would be consistent with the \cors profile just discussed
(MJD 54161).\\
Fig. 10 (7A) (22/3/07) MJD 54181.  The profile is still asymmetrical near the centre as on MJD 54161
but otherwise more symmetrical; except for the suggestion of a depression
in the blue wing centred at $\sim -$15 \kms. \\
Fig. 10 (4B) (16/4/07) MJD 54206. There is now a very clear extra dip on the blue side. The plot can be fitted assuming two blended profiles. These have the following characteristics:
 RV/EW/\vsini : Compt. 1, 19.4/4.76/23.5; Compt 2 $-$23.3/1.0/28.0\\
This profile thus suggests material moving outwards from the star at 
$\sim $43 \kms.\\
Fig. 10 (5B) (30/4/07) MJD 54220.
The profile is again asymmetrical but shows distinct differences from the profile
14 days earlier.\\
MJD 54224. RL10 write ``Spectrum similar to typical maximum".
%MJD 54206. 
%The major blue dip is not seen, though the profile is depressed near
%$-$5 \kms.  and is now more extended to the red.
Fig. 10 (6B) (8/5/07) MJD 54228.  
The profile is now symmetrical or nearly so, in contrast  with the results on
MJD 54220 and MJD 54206, but in agreement with the spectrum in 
 RL10 for MJD 54224. \\
Fig. 10 (8B) (19/5/07) MJD 54239. Profile wide and reasonably symmetrical.\\
Fig. 10 (9B) (23/5/07) MJD54243. Profile wide and reasonably symmetrical. \\
MJD 54243. RL10  write ``Lines much broader than on JD4224 but no change in EW". This is
entirely consistent with the trends of \vsinis and EW illustrated in Fig. 10 and
the various profiles discussed above. The text of RL10 seems to imply some line doubling around this time, though details are not given.\\
Fig. 10 (10B) (30/5/07) MJD 54250.
The profile is symmetrical though there is possibly a small dip at $\sim$--35 \kmss 
on the blue wing.\\
Fig. 10 (6C) (20/6/07) MJD 54272. The profile is now very asymmetrical, with major additional
absorption to the blue, possibly extending out to {\it almost} --100 \kmss from
the central dip.\\
Fig. 10 (7C) (26/6/07) MJD 54278. Profile still shows extra absorption to the blue though not now
so prominent.\\
%The plot shows a
%reduction as the sum of two profiles with a difference in radial velocity of
%38 \kms .\\
%A DESCRIPTION OF 2007 JULY 7 (ON MONTAGE) WOULD BE USEFUL HERE TO COMPARE
%WITH THE FOLLOWING
MJD 54288. RL10  write, ``Line doubling. Neutral lines with blue shifted components
and ionized lines with red shifted.''  The shift is $\sim$$\pm$28 \kms.\\
This was just prior to the start of the decline  $\sim$MJD 54291.\\
In the following some approximate visual magnitudes are entered
as estimated from the AAVSO light curve.\\
Fig. 10 (10C) (12/7/07) MJD54294. The profile is symmetrical but now extremely wide
(%see Fig.~--.
AAVSO $\sim6.6$ mag).\\
Fig. 10 (3D) (18/7/70) MJD 54300. The profile is still extremely wide, but more V-shaped 
than the fitted profile.  That might be due to an additional narrow component
affecting the core. AAVSO $\sim7.0$ mag.\\
Fig. 10 (4D) (19/7/07) MJD 54301.  The next night, with better integration.  The profile is
still very wide and shows `wavy' structure about the fitted profile.\\
% A
%second plot shows a fit with two components.  The narrow component is
%consistent with that suspected on MJD 54300.\\
Fig. 10 (5D) (22/7/07) MJD 54304. The profile is still very wide and with structure.
AAVSO $\sim7.2$ mag. \\
Fig. 10 (9D) (4/8/07) MJD 54317. The profile has now gone into emission.
 AAVSO $\sim10.5$ mag.\\
Fig.10 (10D) (6/8/07) MJD 54319.The profile as MJD54317. AAVSO = $\sim10.9$ mag.

\subsubsection{Preliminary discussion}
The above results show that a major `disturbance' was present in the
atmosphere of R~CrB for at least 130 days prior to the 2007 RCB-type decline.
During most of that interval the AAVSO light-curve shows that there were no
large variations in the visual brightness of the star.  It is clear that
though the nucleation of soot or other particles in the outflow from the
star is of considerable interest, it is a by-product of phenomena in the
stellar atmosphere which developed over a significant period.

The \cors data cover periods immediately prior to two other RCB-type
declines. Those declines began about MJD 51870 and MJD 52680 and were
relatively minor events, the star returning to maximum after approximately
100 days.  At the time of writing (September 2015) the 2007 decline has continued for over
3000 days.  It was also deeper than the two other events.  As can be seen
from Fig.~\ref{coravel}, no variations in EW or \vsinis of sizes comparable to those
observed in the period MJD 54160 to MJD 54300 were seen during those earlier declines. %periods.
This may suggest that the  
extent of a decline is related to the level of prior activity in the
stellar atmosphere.

 The \cors profiles are averages from hundreds of lines and depend on
the fit of the stellar spectrum to the standard that is represented by the
cross-correlating mask.  Whilst caution is necessary in interpreting the
profiles, it is encouraging to see that where high-resolution spectra are
available they generally support a straightforward interpretation. Thus a
high \vsinis is seen when RL see widened spectral lines, and when profile
asymmetries and displaced components are seen, the spectra show high/low-
velocity components.  It seems safe therefore to attribute the widenings,
asymmetries and additional components in the profiles primarily to velocity
effects in the stellar atmosphere, as already noted. It has been
suggested that the puffs of soot which cause the RCB-type declines ultimately
derive from large convective elements in the stellar atmosphere
for which a rough estimate suggested a time-scale of about 40 days
(Feast 1996). If that is so, the variations in the line profile can be
attributed to the combined effects of large turbulent convective
elements.  Some of them are seen individually as displaced velocity
components, whilst other contribute to the general broadening of the profile
(as in normal macro-turbulence).  Whilst we cannot rule out the possibility
that some of the displaced components might be the result of ejection or
infall of complete shells, the rapid changes in profile, on a time scale of a
few days, may be easier to understand as due to turbulent elements.

Since the velocity of escape from the star is roughly 50 \kms, 
the velocities of the displaced components indicate that matter
is being raised to considerable distances above the main stellar atmosphere.

During the period just prior to the 2007 minimum and whilst the star was
still near maximum light, the \cors radial velocities show clear evidence
of a variation with a `period' of 40 to 45 days (see \S3.4.3), but it is of
smaller amplitude than variations on the same time-scale elsewhere in the
series.  The mean velocity at this epoch seems to be slightly negative (by a
few \kms ) in comparison with the mean over a much longer interval (see
\S5). That shows that, although there is evidence of quite high
velocities outwards from the star at that time, the velocity integrated
over the visible hemisphere is showing only a mild general outward motion.

\subsection{Star at maximum in 2006}
{\it Coravel} observations were obtained on a nearly nightly basis for a long
interval in 2006 when the star was away from any major decline.  In the present
section we discuss the observations from MJD 53909 to 53960.
% where the
%detailed \cors profiles have been recorded and are illustrated in Figs.
%--.  
The general trends of radial velocity, EW and \vsinis at that time can
be seen from Fig.~\ref{coravel}.

The radial velocity goes through a rather smooth
cycle of $\sim40$ days with an amplitude of about $5 \rm \, km s^{-1}$.  \vsinis shows a
comparatively large variation (compared to much of the total record) and
seems to be in phase with the radial velocities.  The EW has a rather small
variation which may also be in phase with the radial velocities.  
%The\cors profiles are shown in Figs.~--. 
The following is a brief description
of representative \cors profiles.\\
%NOTE: not all observed profiles are listed here.\\
MJD 53909. Symmetrical, wide.\\
MJD 53914. Going asymmetrical.\\
MJD 53915. Asymmetry very obvious.\\
MJD 53917. ditto\\
MJD 53920. Very asymmetrical. It seems possible that the line might be fitted
with two components, the main one having a minimum to the blue of
the fit shown and the other in the red wing at $\sim$40~\kms.\\ 
MJD 53926. Less asymmetrical.\\
MJD 53928. Component in red wing?\\
MJD 53931. Nearly symmetrical.\\
MJD 53937. Possibly an extra component in the blue wing.\\
MJD 53945. Similar structure to MJD 53920.\\
MJD 53954. Quite asymmetrical.\\
MJD 53955. Very asymmetrical.\\
MJD 53957. Possible component in blue wing and also in red wing.\\  
MJD 53957. Wider scan, same night. Clear component 
$\sim$40 \kmss to blue of main dip.\\
MJD 53961. Component in red wing at $\it \Delta V \sim$+25 \kms .

The above series shows that, at maximum light, changes in the profile shape,
including asymmetries and the appearance of additional components, can take
place on a time scale of a few days.

High-resolution spectroscopy on an almost daily basis is necessary to
  understand these changes in detail. However, in view of the previous
  discussion, the simplest hypothesis is to assume that we are seeing the
  effects of changes due to large turbulent convective cells rising and
  falling in the atmosphere of the star.  Whilst this activity is not as
  extreme as that seen immediately before the major decline of 2007, it is
  of the same general nature.

\subsection{A further example of profile changes}
Fig.~\ref{coravel} shows that from about MJD 52800 to 52900 the radial-velocity
amplitude was large and showed a `periodicity' of about 40 days.  \vsinis
shows large variations which are in phase with the radial velocities.  The
EWs show only a small change.  They may be in anti-phase with the velocities.
Fig.~10 (1C) (9/8/03) shows the profile on MJD 52861. Though symmetrical,
the width is at the maximum value seen until
the 2007 data discussed above. Only seven days later (MJD 52868) the profile is quite
asymmetrical (Fig. 10 (2A) (16/8/03) MJD 52868 EW = 4.0: \vsini = 23). 
   
This is the maximum degree of asymmetry seen in
traces other than that observed in the lead-up to the great decline of 2007.  Skewness in this
sense (`blended to red', i.e. line more extended to the red) is seen more often than in the reverse sense, and in normal (full-brightness) times `blending to blue' is never as conspicuous
as this.

\begin{figure}
\vspace*{-4.0 cm}
\begin{center}
 \includegraphics[width=3.3in]{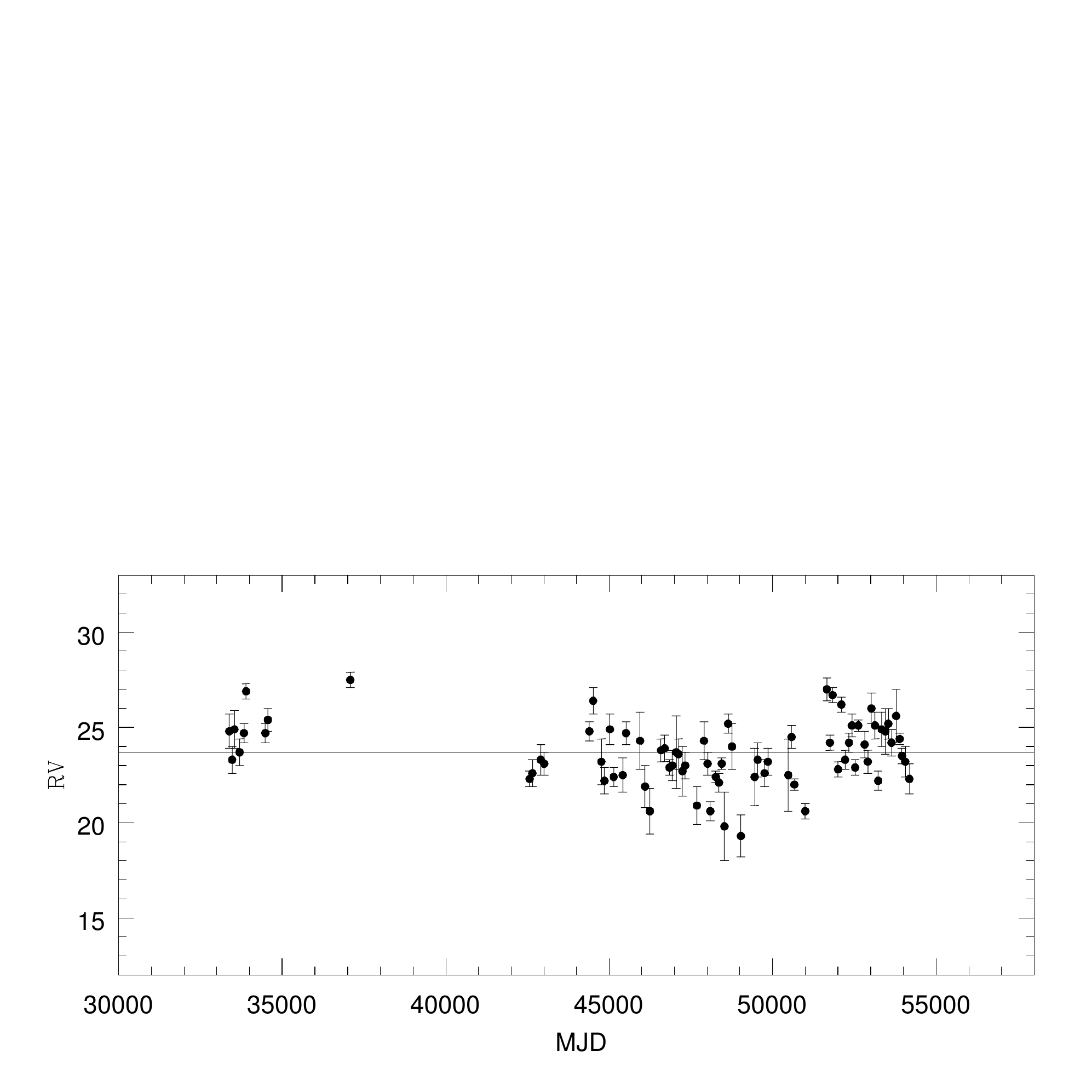} 
% \vspace*{-1.0 cm}
 \caption{Radial velocities, binned as 100 day means, as a function of MJD.}
   \label{ma}
\end{center}
\end{figure}

\section{Long-Term Radial-Velocity Behaviour}

In Fig.~\ref{ma} the radial velocities used in the wavelet analysis  have been 
grouped into 100-day bins and plotted against  MJD. There is an indication
of a variation on a time scale of about ten thousand days. 
This might suggest that R~CrB is a member of a spectroscopic binary.
If so, the two components
would be well separated and would not interact unless the orbit were highly
eccentric.

Most RCB stars show irregular variations in the flux from the surrounding
dust shell on a time-scale of several thousand days (Feast et al.~1997).
The causes of such variations are not known, but are presumed to be
associated with changes in the dust-production mechanism.  Magnetic activity
has been suggested by analogy with the solar cycle (Clayton et al.~1993). If
the primary cause of mass ejection is turbulent eddies, as discussed below,
that may be a plausible mechanism, though it clearly needs
confirmation.

Unlike all other well-studied RCB stars, in which the 
long-term dust flux variations are clearly irregular, there is some evidence for
semi-regular variations in the case of R~CrB with a period of about 1260 days
(Feast et al.~1997, fig.~19). Plots of the 100-day means phased on that period show no 
evidence for a similar variation in the radial velocities.

\section{Small scale variability and RCB type declines}

The observations reported here show that the radial velocity of
R~CrB frequently has variations of the order of 10~\kmss on a time-scale of
around 40 days.  However, the long time span covered by the
observations shows clearly that there is no coherent periodic variation in
the data.  The shape and amplitude of the velocity curve is very variable.
That is consistent with the data in the much shorter previous runs of 
radial-velocity observations and also of photometry.  

This time-scale in the light and velocity variations of R~CrB
has often been claimed as evidence of radial pulsation, which has been
  regarded as providing the basic mechanism for the mass ejection
associated with RCB events.  The observations presented here, however, show
that if any strict periodicity exists  in the radial velocities it must be of a much lower amplitude
than the variations that have been documented. Furthermore the velocities involved are too low
to eject material to cool regions where dust could form.

The data from the \cors
series show the complex atmospheric effects which can occur prior to 
RCB-type declines as well as between declines.
These observations are consistent with the presence of large-scale turbulent
elements in the stellar atmosphere, in some of which 
matter
attains an outward velocity high
enough to carry it sufficiently far from the star to allow  the formation of the dust
puffs that obscure the star in its characteristic declines. A
rough estimate (Feast 1996) indicates that a time-scale of the order of a
month, consistent with that of the observed variability, would be associated
with those turbulent elements.  

It is not known whether the 40--50-day cycles in light and velocity, 
when they occur, are global or not. However,
the hypothesis that the light and velocity
variations are related to the integrated effects of a small number of giant
turbulent convective cells over the visible side of the star, rather
than a global pulsation,
is consistent with the view that the ejection of
dust puffs is related to turbulent elements.

It has been suggested (Crause et al. 2007) that the deep minima of R~CrB
repeat regularly as multiples of $42.97 \pm 0.03$ days, and that has been
claimed as evidence for a relationship between a pulsation 
of that period and the RCB events.
The analysis of Crause et al. is based on 
AAVSO visual observations of 18 declines over 9493
days. The period derived by Gorynya et al. (1992) was adopted from the range
of periods in the literature as the one best fitting the observations.  The
very small uncertainty in the adopted period is required if the declines are to
be kept in even approximate phase.
The Gorynya et al. period is from 1990 only (it was not seen by them in
1991, see above) and must have a far greater uncertainty than that
(see also the discussion in \S3.3.3).  

The Crause et al. data are relatively sparse.  About half the observations refer
to multiple declines, when the star undergoes further dimming before it has
recovered from an earlier decline or where two declines are close together.
In such cases we might expect (as in the light curve at maximum) that some approximate
`keeping of phase' would occur if the effect is due to a single convective
element or several working together.  The scatter is also significant, the
range in differences from the predicted decline is from $+13$ to $-13$ days
and appears to rule out any precise correspondence of
declines with the adopted period. 

We showed in section 3.3 that there is no sign of a coherent period with the Crause et al. value. 
Table 4  shows this in a somewhat different way. It lists 
dates of well-defined radial-velocity minima in the \cors series (see Fig.~\ref{coravel}).
 
Table 4 also lists the nearest date given by the ephemeris
of Crause et al. Since that does not necessarily refer to minimum velocity
one might expect a constant offset, if the star varied with the adopted period.
Instead one sees from the table that all possible offsets are found, 
further evidence that the observations give no support to that period. 
Alternatively one can compare the first well-defined velocity minimum
in the older Griffin series (MJD 42584) with a particularly well defined
minimum towards the end of the \cors series (MJD 53178),
an interval of 10594 days. With a
period of 42.97 days those minima are exactly out of phase ($\Delta phase = 0.46$).
Changing the period to bring these two minima into phase would destroy even a very
approximate correspondence
with RCB-type declines.

It seems possible that the effect reported by Crause et al.
 is a chance phenomenon connected with  the small number of independent declines and the fact that the ``period'' was chosen 
from the many in the literature as the one fitting the decline data best.

\setcounter{table}{3}
\begin{table}
\centering
\caption{MJD of velocity minima in the Coravel series}
\begin{tabular}{ccr}
\hline
Observed&Predicted&$\Delta$\\
\hline
51594\rlap{:}&51615&--21\\
51977&51958&+19\\
52435&52431&+4\\
52477&52474&+3\\
52825&52818&+7\\
52871&52861&+10\\
53178&53161&+17\\
53254&53247&+7\\
53521&53505&+16\\
53622&53634&--12\\
53926&53935&--9\\
54215\rlap{:}&54236&--21\\
\hline
\end{tabular}
\end{table}

It is interesting to compare R~CrB with RY Sgr which is also a member of the
RCB class. RY~Sgr has considerably larger variations in light and radial
velocity ($\sim$0.5 mag, $\sim$25~\kms ) than R~CrB.  Infrared data when the
star is in a deep obscuration phase strongly suggest that the variations are
global and are due to radial pulsations with a period of about 38.6
days\footnote{In principle it would be possible to carry out a similar test
in the case of R~CrB.}.  The light and radial-velocity variations do not
repeat exactly, indicating some irregular variations superposed on the
pulsations (see Alexander et al. 1972, their figs. 3 to 8).  Although RY~Sgr shows
much clearer and larger-amplitude pulsations than any other known RCB star
it does not stand out as a dust producer.  That has long been seen as a
problem in attributing dust production to pulsational ejection of matter.
In view of the discussion above it seems reasonable to postulate that, as in
R~CrB, the primary ejection mechanism is associated with large atmospheric
turbulent elements, though certain phases in the pulsation cycle of RY Sgr might be
 favoured\footnote{In the case of RY Sgr, at least,
a different problem occurs in the phasing of the light declines.  Evidently
if decline occurs between pulsation maximum and minimum it will tend to
appear to start near maximum.  If it starts between minimum and maximum it
will appear to start nearer maximum than was actually the case, owing to the
increase in light from the pulsation.  Any statistics of phasing of declines
would need to take this into account.}.

Clearly it is not possible to rule out the hypothesis that R~CrB is
undergoing some regular pulsations of very low amplitude, but the current
data do not require that hypothesis and such a low amplitude could not
provide the velocities necessary to eject matter from the star. 

\section {Final Conclusions}
The observations reported in this paper are not consistent with 
a coherent pulsational model for R~CrB.
At certain times and especially before the great decline of 2007, there is
evidence that the atmosphere is disturbed, with  high-velocity
components in the line profiles. We suggest that they are connected with large turbulent
elements in the stellar atmosphere and result in mass ejection to sufficient
distance from the star that solid particles can be formed leading
to an RCB-type decline. Once solids are formed, radiation
pressure will drive the dust and entrained gas from the star at velocities
of $\sim$200~\kms\ or more, as is observed. A few large turbulent, convective cells
with a characteristic turn-over time of
$\sim$40 days, appear to provide a suitable explanation of the
light and radial-velocity variations seen in R~CrB outside RCB-type declines.
\section*{Acknowledgments}
MWF and PAW gratefully acknowledge the receipt of research grants from the
National Research Foundation (NRF) of South Africa. MWF thanks Dr L Balona
for some help with the data for the early years.

\newpage
\appendix
\section{Tables of R~CrB radial velocities}
\label{appA}

\subsection{Table A1}

The (heliocentric) radial velocities, Vel (\kmss), of R~CrB discussed in section~\ref{Sec_herbig}, with the MJD on which they were obtained; note: ``0.5"  indicates an underexposed
spectrogram. 
\begin{center}
\scriptsize
\tablefirsthead{  
\hline
MJD  &  Vel &  note     \\
\hline}
\noindent
%\topcaption{The (heliocentric) radial velocities of R~CrB discussed in section~\ref{Sec_herbig}}
\begin{supertabular}{lcc}
%\tablecaption{The (heliocentric) radial velocities of R~CrB discussed in section~\ref{Sec_herbig}}
\tablehead{
\hline
MJD     &   Vel &  note     \\
%\multicolumn{3}{|l|}{\small\sl continued from previous column}\\
\hline}
\noindent
\tabletail{
\hline
\multicolumn{3}{|r|}{\small\sl continued on next column}\\
\hline}
\tablelasttail{\hline}
   33321.53 &    23.0&      \\
   33383.40 &    27.9&       \\
   33390.44 &    24.6&       \\
   33397.33 &    23.2&       \\
   33413.23 &    25.2&       \\
   33424.23 &    27.2&       \\
   33433.21 &    21.3&       \\
   33446.24 &    22.5&       \\
   33451.21 &    24.2&       \\
   33452.19 &    24.3&       \\
   33458.31 &    25.3&       \\
   33467.27 &    24.8&       \\
   33476.20 &    24.0&       \\
   33483.21 &    25.2&       \\
   33496.20 &    24.6&   0.5 \\
   33506.19 &    19.7&       \\
   33513.18 &    18.9&       \\
   33517.17 &    20.7&       \\
   33525.16 &    24.2&       \\
   33535.16 &    27.1&       \\
   33536.14 &    27.5&   0.5 \\
   33546.14 &    22.7&       \\
   33553.14 &    23.2&       \\
   33653.56 &    25.6&       \\
   33671.51 &    24.7&       \\
   33678.51 &    24.3&       \\
   33716.43 &    21.4&       \\
   33720.43 &    21.5&       \\
   33729.35 &    24.5&       \\
   33775.31 &    25.4&       \\
   33776.41 &    25.8&   0.5 \\
   33786.47 &    20.7&       \\
   33805.21 &    25.7&       \\
   33808.36 &    25.2&       \\
   33814.28 &    23.2&       \\
   33818.31 &    21.6&       \\
   33822.31 &    23.3&       \\
   33825.26 &    22.6&       \\
   33830.27 &    23.9&   0.5 \\
   33832.25 &    24.4&   0.5 \\
   33836.24 &    23.6&       \\
   33839.25 &    23.5&       \\
   33845.19 &    23.6&       \\
   33850.18 &    24.9&       \\
   33854.21 &    26.0&       \\
   33858.19 &    28.2&       \\
   33862.30 &    27.1&       \\
   33865.19 &    26.6&       \\
   33874.17 &    28.4&       \\
   33878.17 &    27.0&       \\
   33882.22 &    25.9&       \\
   33885.17 &    25.0&   0.5 \\
   33892.17 &    26.7&       \\
   33895.16 &    27.5&       \\
   33900.16 &    27.5&       \\
   33916.16 &    28.5&       \\
   34433.48 &    26.6&       \\
   34450.41 &    28.0&       \\
   34466.50 &    23.9&       \\
   34467.45 &    23.1&       \\
   34470.50 &    22.4&       \\
   34478.50 &    22.5&       \\
   34482.51 &    23.6&       \\
   34488.46 &    24.5&       \\
   34492.44 &    26.5&       \\
   34502.35 &    25.9&       \\
   34509.41 &    24.3&       \\
   34525.29 &    25.4&   0.5 \\
   34542.31 &    26.0&       \\
   34551.29 &    26.6&       \\
   34559.25 &    25.7&       \\
   34573.35 &    25.5&       \\
   34584.27 &    23.1&       \\
\end{supertabular}
%\end{table}
\end{center}

\subsection{Table A2}
MJD and radial velocities, Vel (\kms), of R~CrB by R. Griffin using various telescopes coded as follows: 218 Palomar,
303 OHP,
123 ESO,
219 DAO,
202 Original Cambridge,
302 Cambridge Coravel,
Uncoded Cambridge Coravel. An o in the note column indicates that the observation was made during a faint phase.
\begin{center}
\scriptsize
\tablefirsthead{  
\hline
MJD     &    note &  Vel &  code     \\
\hline}
\begin{supertabular}{cccc}
\tablehead{
\hline
MJD     &    note &  Vel &  code     \\
%\multicolumn{5}{|l|}{\small\sl continued from previous column}\\
\hline}
\tabletail{
\hline
\multicolumn{4}{|r|}{\small\sl continued on next column}\\
\hline}
\tablelasttail{\hline}
   39970.04&   &  26.4&       202   \\
   40328.12&   &  26.0&       202   \\
   40671.17&   &  29.7&       202   \\
   40676.17&   &  24.8&       202   \\
   40787.89&   &  24.2&       202   \\
   40797.90&   &  25.9&       202   \\
   40996.17&   &  19.8&       202   \\
   41150.89&   &  26.9&       202   \\
   41514.93&   &  25.3&       202   \\
   41798.11&   &  25.2&       202   \\
   42152.09& o &  27.7&       202   \\
   42200.43& o &  23.6&       218   \\
   42249.95&   &  26.6&       202   \\
   42280.86&   &  19.4&       202   \\
   42312.79&   &  28.7&       202   \\
   42510.07&   &  22.5&       202   \\
   42512.12&   &  23.4&       202   \\
   42522.08&   &  21.6&       202   \\
   42532.01&   &  22.6&       202   \\
   42535.06&   &  24.3&       202   \\
   42536.99&   &  23.6&       202   \\
   42554.49&   &  26.9&       218   \\
   42563.02&   &  24.8&       202   \\
   42568.01&   &  25.3&       202   \\
   42568.96&   &  22.3&       202   \\
   42569.94&   &  25.2&       202   \\
   42570.98&   &  23.2&       202   \\
   42571.90&   &  23.2&       202   \\
   42572.90&   &  21.2&       202   \\
   42574.90&   &  21.4&       202   \\
   42575.96&   &  21.6&       202   \\
   42579.04&   &  21.8&       202   \\
   42580.98&   &  19.8&       202   \\
   42583.97&   &  18.5&       202   \\
   42584.94&   &  20.0&       202   \\
   42585.89&   &  19.0&       202   \\
   42586.89&   &  20.2&       202   \\
   42588.01&   &  21.4&       202   \\
   42588.90&   &  20.9&       202   \\
   42592.02&   &  20.8&       202   \\
   42592.90&   &  22.7&       202   \\
   42595.89&   &  25.0&       202   \\
   42626.89&   &  27.4&       202   \\
   42630.86&   &  24.8&       202   \\
   42637.85&   &  22.4&       202   \\
   42648.87&   &  19.0&       202   \\
   42649.86&   &  21.5&       202   \\
   42650.85&   &  20.5&       202   \\
   42652.84&   &  21.8&       202   \\
   42660.84&   &  23.1&       202   \\
   42670.82&   &  25.6&       202   \\
   42675.85&   &  21.4&       202   \\
   42676.78&   &  23.4&       202   \\
   42685.84&   &  19.8&       202   \\
   42703.75& o &  16.8&       202   \\
   42708.74& o &  16.3&       202   \\
   42801.25& o &  26.6&       202   \\
   42838.16& o &  27.6&       202   \\
   42872.07&   &  21.6&       202   \\
   42887.94&   &  25.6&       202   \\
   42891.05&   &  22.7&       202   \\
   42902.06&   &  22.2&       202   \\
   42934.93&   &  21.2&       202   \\
   42935.97&   &  23.6&       202   \\
   42941.98&   &  26.4&       202   \\
   42980.90&   &  21.9&       202   \\
   42985.90&   &  22.7&       202   \\
  42987.88&  & 24.4&         202 \\
  42990.87 &  & 23.9&        202 \\
  43005.84&  & 21.4&        202 \\
   43006.85&   &  20.8&       202   \\
   43009.84&   &  21.0&       202   \\
   43039.82&   &  26.6&       202   \\
   43047.78&   &  24.5&       202   \\
   43055.78&   &  23.3&       202   \\
   43178.21&   &  25.4&       202   \\
   43180.18&   &  23.7&       202   \\
   43593.09&   &  28.0&       202   \\
   43598.10&   &  28.1&       202   \\
   43679.97&   &  26.7&       202   \\
   43736.84&   &  17.8&       202   \\
   43739.84&   &  20.6&       202   \\
   43886.24&   &  21.8&       202   \\
   43929.21&   &  25.4&       202   \\
   43944.14&   &  25.9&       202   \\
   43988.07&   &  26.1&       202   \\
   43996.04&   &  22.2&       202   \\
   44007.02&   &  21.9&       202   \\
   44017.00&   &  23.8&       202   \\
   44113.83&   &  19.9&       202   \\
   44137.79&   &  19.8&       202   \\
   44232.30&   &  23.4&       202   \\
   44238.23&   &  22.1&       202   \\
   44240.31&   &  21.2&       202   \\
   44292.19&   &  23.2&       202   \\
   44363.00&   &  23.4&       202   \\
   44365.03&   &  23.4&       202   \\
   44368.05&   &  23.9&       202   \\
   44369.98&   &  25.8&       202   \\
   44372.02&   &  23.2&       202   \\
   44374.03&   &  23.9&       202   \\
   44376.04&   &  25.9&       202   \\
   44378.02&   &  27.7&       202   \\
   44389.00&   &  28.4&       202   \\
   44398.03&   &  25.7&       202   \\
   44417.40&   &  21.5&       218   \\
   44434.90&   &  25.5&       202   \\
   44441.91&   &  23.4&       202   \\
   44444.89&   &  22.7&       202   \\
   44447.94&   &  26.9&       202   \\
   44471.87&   &  22.2&       202   \\
   44477.83&   &  25.0&       202   \\
   44482.82&   &  22.8&       202   \\
   44488.84&   &  27.9&       202   \\
   44503.80&   &  30.2&       202   \\
   44506.80&   &  30.1&       202   \\
   44516.78&   &  26.5&       202   \\
   44521.77&   &  25.6&       202   \\
   44526.80&   &  27.3&       202   \\
   44545.73&   &  28.8&       202   \\
   44559.71&   &  26.0&       202   \\
   44561.71&   &  24.7&       202   \\
   44580.27&   &  22.8&       202   \\
   44621.29&   &  22.6&       202   \\
   44676.15&   &  20.1&       202   \\
   44712.08&   &  20.8&       202   \\
   44727.02&   &  17.6&       202   \\
   44729.06&   &  16.9&       202   \\
   44743.43&   &  29.4&       218   \\
   44748.91&   &  28.9&       202   \\
   44753.97&   &  28.8&       202   \\
   44776.90&   &  21.6&       202   \\
   44783.96&   &  20.7&       202   \\
   44789.94&   &  22.3&       202   \\
   44801.92&   &  24.4&       202   \\
   44803.95&   &  24.0&       202   \\
   44811.87&   &  23.4&       202   \\
   44814.87&   &  19.6&       202   \\
   44818.87&   &  20.3&       202   \\
   44826.85&   &  19.3&       202   \\
   44832.86&   &  22.3&       202   \\
   44842.84&   &  26.2&       202   \\
   44859.80&   &  22.2&       202   \\
   44860.82&   &  18.7&       202   \\
   44865.84&   &  20.5&       202   \\
   44868.83&   &  23.2&       202   \\
   44872.83&   &  21.7&       202   \\
   44879.79&   &  24.5&       202   \\
   44889.79&   &  25.7&       202   \\
   44893.76&   &  24.1&       202   \\
   44979.29&   &  24.4&       202   \\
   44990.27&   &  26.2&       202   \\
   45032.18&   &  24.4&       202   \\
   45044.14&   &  27.1&       202   \\
   45074.05&   &  22.4&       202   \\
   45093.04&   &  23.9&       202   \\
   45096.05&   &  23.6&       202   \\
   45100.95&   &  23.6&       202   \\
   45112.99&   &  23.9&       202   \\
   45115.02&   &  25.4&       202   \\
   45118.97&   &  25.6&       202   \\
   45143.02&   &  21.1&       202   \\
   45145.99&   &  23.4&       202   \\
   45148.96&   &  22.7&       202   \\
   45149.96&   &  21.9&       202   \\
   45153.95&   &  20.5&       202   \\
   45157.91&   &  18.9&       202   \\
   45159.95&   &  19.7&       202   \\
   45165.99&   &  21.7&       202   \\
   45169.88&   &  19.1&       202   \\
   45176.88&   &  20.2&       202   \\
   45192.90&   &  25.3&       202   \\
   45212.84&   &  24.0&       202   \\
   45214.83& o &  26.6&       202   \\
   45225.84& o &  29.0&       202   \\
   45230.84& o &  29.7&       202   \\
   45233.79& o &  27.4&       202   \\
   45235.85& o &  26.0&       202   \\
   45269.73&   &  27.4&       202   \\
   45368.58&   &  18.9&       219   \\
   45369.60&   &  18.4&       219   \\
   45383.50&   &  22.1&       219   \\
   45388.18&   &  23.2&       202   \\
   45404.16&   &  22.0&       202   \\
   45408.15&   &  19.8&       202   \\
   45440.07&   &  25.5&       202   \\
   45446.00&   &  27.2&       202   \\
   45448.08&   &  25.9&       202   \\
   45457.95&   &  22.6&       202   \\
   45464.05&   &  21.6&       202   \\
   45470.01&   &  22.6&       202   \\
   45471.04&   &  23.7&       202   \\
   45479.01&   &  27.6&       202   \\
   45491.97&   &  29.3&       202   \\
   45493.98&   &  27.4&       202   \\
   45497.96&   &  23.1&       202   \\
   45499.99&   &  22.7&       202   \\
   45506.90&   &  20.3&       202   \\
   45515.94&   &  23.5&       202   \\
   45517.95&   &  24.4&       202   \\
   45519.94&   &  27.6&       202   \\
   45527.90&   &  30.3&       202   \\
   45536.88&   &  25.1&       202   \\
   45544.88&   &  22.6&       202   \\
   45548.88&   &  23.1&       202   \\
   45549.89&   &  23.5&       202   \\
   45557.87&   &  27.4&       202   \\
   45564.95&   &  22.9&       202   \\
   45566.92&   &  22.8&       202   \\
   45804.11& o &  27.2&       202   \\
   45807.10& o &  23.9&       202   \\
   45814.06& o &  29.5&       202   \\
   45816.07& o &  27.7&       202   \\
   45818.09& o &  26.6&       202   \\
   45829.06& o &  21.8&       202   \\
   45833.03& o &  24.9&       202   \\
   45859.95& o &  25.3&       202   \\
   45892.94& o &  17.0&       202   \\
   45909.88&   &  19.0&       202   \\
   45932.85&   &  21.1&       202   \\
   45942.84&   &  27.7&       202   \\
   45944.87&   &  28.7&       202   \\
   45947.83&   &  28.5&       202   \\
   45965.80&   &  24.4&       202   \\
   45972.77&   &  20.6&       202   \\
   46055.30&   &  22.0&       202   \\
   46077.29&   &  24.9&       202   \\
   46089.24&   &  23.9&       202   \\
   46113.56&   &  18.9&       219   \\
   46120.18&   &  18.2&       202   \\
   46135.19&   &  23.8&       202   \\
   46216.00&   &  25.1&       202   \\
   46217.97&   &  24.5&       202   \\
   46248.95&   &  18.4&       202   \\
   46249.93&   &  16.8&       202   \\
   46252.93&   &  18.6&       202   \\
   46263.04&   &  19.7&       202   \\
   46266.97&   &  20.8&       202   \\
   46314.86& o &  22.7&       202   \\
   46323.87& o &  24.3&       202   \\
   46455.24& o &  21.8&       202   \\
   46488.20& o &  14.6&       202   \\
   46495.14& o &  12.9&       202   \\
   46525.11& o &  24.5&       303   \\
   46530.16& o &  21.3&       303   \\
   46556.05&   &  23.3&       202   \\
   46563.05&   &  25.0&       202   \\
   46564.04&   &  25.9&       202   \\
   46566.03&   &  25.9&       202   \\
   46567.01&   &  25.8&       202   \\
   46569.03&   &  26.5&       202   \\
   46576.01&   &  25.2&       202   \\
   46577.01&   &  23.4&       202   \\
   46579.00&   &  24.9&       202   \\
   46580.92&   &  23.5&       202   \\
   46584.97&   &  24.8&       202   \\
   46591.97&   &  18.2&       202   \\
   46592.96&   &  20.7&       202   \\
   46593.92&   &  19.0&       202   \\
   46595.96&   &  19.3&       202   \\
   46610.95&   &  22.9&       202   \\
   46614.97&   &  24.7&       202   \\
   46634.89&   &  25.8&       202   \\
   46648.89&   &  26.5&       202   \\
   46658.95&   &  25.3&       202   \\
   46664.83&   &  24.2&       303   \\
   46665.87&   &  22.8&       303   \\
   46666.87&   &  23.6&       303   \\
   46667.85&   &  24.2&       303   \\
   46670.87&   &  22.8&       303   \\
   46689.82&   &  28.4&       202   \\
   46691.81&   &  29.2&       202   \\
   46698.79&   &  26.7&       202   \\
   46701.85&   &  27.5&       202   \\
   46707.77&   &  21.7&       202   \\
   46714.76&   &  22.1&       202   \\
   46740.73&   &  21.8&       202   \\
   46742.72&   &  20.8&       202   \\
   46746.72&   &  21.5&       202   \\
   46749.71&   &  20.5&       202   \\
   46776.29&   &  24.3&       202   \\
   46826.24&   &  20.7&       202   \\
   46845.23&   &  21.9&       202   \\
   46847.21&   &  23.1&       202   \\
   46850.24&   &  25.5&       202   \\
   46855.21&   &  22.4&       303   \\
   46857.23&   &  22.4&       303   \\
   46859.15&   &  21.8&       303   \\
   46874.11&   &  24.1&       202   \\
   46876.17&   &  23.2&       202   \\
   46880.13&   &  23.8&       202   \\
   46912.99&   &  19.1&       202   \\
   46917.06&   &  20.2&       202   \\
   46920.03&   &  22.7&       202   \\
   46923.05&   &  25.8&       202   \\
   46924.02&   &  25.9&       202   \\
   46931.02&   &  24.1&       202   \\
   46934.98&   &  22.9&       202   \\
   46947.00&   &  19.2&       202   \\
   46949.98&   &  20.0&       202   \\
   46968.95&   &  23.2&       202   \\
   46970.94&   &  23.8&       202   \\
   46979.95&   &  28.3&       202   \\
   46981.97&   &  29.0&       202   \\
   47052.84&   &  18.9&       202   \\
   47062.84&   &  19.8&       202   \\
   47074.78&   &  24.8&       202   \\
   47080.74&   &  25.8&       303   \\
   47085.73&   &  23.7&       303   \\
   47089.74&   &  22.7&       202   \\
   47090.75&   &  23.6&       202   \\
   47092.74&   &  22.4&       202   \\
   47096.74&   &  20.3&       202   \\
   47137.27&   &  24.8&       202   \\
   47139.29&   &  23.5&       202   \\
   47151.29&   &  22.3&       202   \\
   47168.24&   &  23.6&       202   \\
   47183.61&   &  29.6&       219   \\
   47190.57&   &  29.9&       219   \\
   47192.62&   &  30.2&       219   \\
   47231.22&   &  19.8&       303   \\
   47232.14&   &  18.9&       303   \\
   47234.17&   &  19.4&       303   \\
   47242.18&   &  22.9&       202   \\
   47260.95&   &  22.3&       202   \\
   47264.09&   &  23.2&       202   \\
   47277.08&   &  20.1&       202   \\
   47287.95&   &  19.9&       202   \\
   47300.98&   &  21.4&       202   \\
   47302.97&   &  23.6&       202   \\
   47307.96&   &  22.6&       202   \\
   47309.91&   &  21.5&       202   \\
   47314.99&   &  21.2&       202   \\
   47318.96&   &  21.3&       202   \\
   47324.96&   &  19.3&       202   \\
   47334.93&   &  22.8&       202   \\
   47349.05&   &  26.0&       202   \\
   47353.93&   &  27.1&       202   \\
   47357.92&   &  26.0&       202   \\
   47360.88& o &  24.0&       202   \\
   47368.89& o &  30.4&       202   \\
   47467.75& o &  29.9&       303   \\
   47470.74& o &  31.5&       303   \\
   47568.23& o &  32.4&       202   \\
   47580.42& o &  22.2&       123   \\
   47582.41& o &  22.5&       123   \\
   47603.10& o &  28.7&       202   \\
   47612.15& o &  24.1&       303   \\
   47614.16& o &  23.5&       303   \\
   47644.10&   &  18.3&       303   \\
   47646.13&   &  18.7&       303   \\
   47649.09&   &  20.0&       303   \\
   47665.92&   &  25.9&       202   \\
   47673.00&   &  23.2&       202   \\
   47676.99&   &  18.7&       202   \\
   47679.00&   &  16.8&       202   \\
   47681.96&   &  15.3&       202   \\
   47689.93&   &  16.0&       202   \\
   47695.95&   &  18.0&       202   \\
   47700.94&   &  22.4&       202   \\
   47702.94&   &  22.0&       202   \\
   47710.00&   &  27.1&       202   \\
   47711.94&   &  27.1&       202   \\
   47718.94&   &  24.7&       202   \\
   47731.91& o &  21.3&       202   \\
   47741.91& o &  24.1&       202   \\
   47743.87& o &  23.3&       202   \\
   47750.90& o &  32.0&       202   \\
   47765.84& o &  30.5&       202   \\
   47775.87& o &  27.0&       202   \\
   47816.75& o &  23.8&       202   \\
   47826.75& o &  27.8&       303   \\
   47828.76& o &  29.3&       303   \\
   47830.72& o &  29.2&       303   \\
   47831.72& o &  28.2&       303   \\
   47834.73& o &  26.2&       202   \\
   47841.73&   &  24.1&       202   \\
   47846.71&   &  20.2&       202   \\
   47918.25&   &  26.6&       303   \\
   47922.21&   &  23.4&       303   \\
   47934.41&   &  25.4&       123   \\
   47936.41&   &  26.1&       123   \\
   47963.47&   &  21.4&       219   \\
   47977.09&   &  28.8&       202   \\
   47979.11&   &  28.1&       202   \\
   47986.08&   &  25.2&       202   \\
   47999.06&   &  20.2&       202   \\
   48010.08&   &  22.2&       202   \\
   48012.03&   &  22.5&       202   \\
   48014.97&   &  25.5&       202   \\
   48028.06&   &  26.0&       202   \\
   48035.93&   &  19.7&       202   \\
   48038.00&   &  18.1&       202   \\
   48061.96&   &  28.1&       202   \\
   48074.99&   &  22.6&       202   \\
   48077.91&   &  19.8&       202   \\
   48080.90&   &  19.8&       202   \\
   48083.96&   &  17.8&       202   \\
   48085.94&   &  18.2&       202   \\
   48090.92&   &  21.2&       202   \\
   48096.95&   &  24.1&       202   \\
   48170.83&   &  25.2&       202   \\
   48189.73&   &  21.7&       202   \\
   48229.28&   &  20.6&       202   \\
   48282.19&   &  22.8&       303   \\
   48283.20&   &  23.0&       303   \\
   48284.22&   &  22.6&       303   \\
   48285.21&   &  22.9&       303   \\
   48286.21&   &  23.1&       303   \\
   48290.21&   &  23.8&       303   \\
   48292.23&   &  22.4&       303   \\
   48349.05&   &  24.4&       202   \\
   48385.04&   &  19.9&       202   \\
   48399.04&   &  25.6&       202   \\
   48401.01&   &  23.1&       202   \\
   48418.00&   &  24.5&       202   \\
   48420.99&   &  23.9&       202   \\
   48452.90&   &  22.7&       202   \\
   48455.92&   &  21.5&       202   \\
   48460.97&   &  20.0&       202   \\
   48465.88&   &  20.8&       202   \\
   48557.73&   &  23.9&       303   \\
   48561.72&   &  24.1&       303   \\
   48607.25&   &  27.5&       303   \\
   48609.25&   &  26.1&       303   \\
   48636.25&   &  24.4&       303   \\
   48639.26&   &  24.2&       303   \\
   48641.26&   &  24.5&       303   \\
   48678.56&   &  25.1&       219   \\
   48680.53&   &  24.3&       219   \\
   48734.11&   &  26.5&       303   \\
   48736.11&   &  24.0&       303   \\
   48738.11&   &  21.6&       303   \\
   48739.12&   &  21.1&       303   \\
   48741.12&   &  19.7&       303   \\
   48743.12&   &  19.4&       303   \\
   48793.05&   &  26.1&       303   \\
   48797.99&   &  29.0&       303   \\
   48800.02&   &  28.2&       303   \\
   48846.90&   &  21.1&       303   \\
   48850.92&   &  21.4&       303   \\
   48974.26&   &  22.4&       303   \\
   48975.26&   &  24.1&       303   \\
   48976.26&   &  23.7&       303   \\
   49029.24&   &  20.7&       303   \\
   49032.23&   &  20.7&       303   \\
   49034.11&   &  19.5&       303   \\
   49065.18&   &  15.2&       303   \\
   49066.20&   &  15.5&       303   \\
   49069.20&   &  15.3&       303   \\
   49071.19&   &  16.2&       303   \\
   49173.95&   &  19.0&       303   \\
   49175.99&   &  18.7&       303   \\
   49178.97&   &  22.4&       303   \\
   49348.26& o &  20.8&       303   \\
   49351.25& o &  21.9&       303   \\
   49360.26& o &  20.2&       303   \\
   49401.17& o &  17.6&       303   \\
   49472.09&   &  23.2&       303   \\
   49563.83&   &  22.8&       303   \\
   49567.91&   &  24.0&       303   \\
   49570.87&   &  25.9&       303   \\
   49697.25&   &  26.6&       303   \\
   49714.25&   &  22.2&       303   \\
   49720.26&   &  20.2&       303   \\
   49727.25&   &  23.3&       303   \\
   49873.05&   &  20.9&       303   \\
   50433.25&   &  17.8&       303   \\
   50442.25&   &  17.9&       303   \\
   50487.22&   &  24.6&             \\
   50508.16&   &  25.5&             \\
   50513.21&   &  26.7&             \\
   50536.15&   &  25.9&             \\
   50538.13&   &  26.2&             \\
   50539.14&   &  25.9&             \\
   50546.10&   &  25.7&             \\
   50548.12&   &  27.1&             \\
   50549.13&   &  27.1&             \\
   50554.11&   &  29.0&             \\
   50563.17&   &  27.8&       303   \\
   50564.12&   &  28.1&       303   \\
   50566.16&   &  27.1&       303   \\
   50569.11&   &  27.8&             \\
   50571.08&   &  27.8&             \\
   50571.96&   &  27.1&             \\
   50573.09&   &  26.0&             \\
   50575.05&   &  24.2&             \\
   50578.05&   &  22.5&             \\
   50579.06&   &  22.0&             \\
   50580.04&   &  22.4&             \\
   50581.08&   &  23.0&             \\
   50609.03&   &  21.8&             \\
   50611.97&   &  20.3&             \\
   50616.04&   &  19.1&             \\
   50617.01&   &  20.4&             \\
   50619.06&   &  21.0&             \\
   50620.03&   &  21.0&             \\
   50630.91&   &  22.8&             \\
   50635.95&   &  23.7&             \\
   50646.98&   &  23.6&       303   \\
   50647.93&   &  22.5&       303   \\
   50649.96&   &  20.8&       303   \\
   50651.86&   &  21.9&       303   \\
   50652.90&   &  21.7&       303   \\
   50653.88&   &  22.1&       303   \\
   50654.93&   &  20.2&       303   \\
   50655.89&   &  21.3&       303   \\
   50656.89&   &  22.4&       303   \\
   50657.85&   &  21.5&       303   \\
   50699.78&   &  21.1&       303   \\
   50701.82&   &  22.1&       303   \\
   50703.77&   &  23.3&       303   \\
   50705.77&   &  23.6&       303   \\
   50804.26&   &  21.8&       303   \\
   50932.09&   &  19.4&       303\\
   50937.10&   &  21.1&       303\\
   51002.00&   &  21.9&       303\\
   51003.99&   &  21.3&       303\\
   51008.95&   &  18.4&       303\\
   51016.91&   &  19.0&       303\\
   51019.94&   &  20.9&       303\\
   51021.91&   &  21.6&       303\\
   51024.88&   &  22.0&       303\\
   51270.47&   &  22.7&       219\\
   51281.49&   &  22.2&       219\\
   51366.33&   &  20.8&       219\\
   51369.32&   &  22.7&       219\\
   51373.30&   &  23.2&       219\\

\end{supertabular}
\end{center}

\subsection{Table A3}

  This details R~CrB  Cambridge Coravel data with recorded profiles;
   o indicates an observation during a  faint phase, while x indicates an observation during the 
   2006 period, when the profiles were abnormal.

\begin{center}
\scriptsize
%\centering
%\begin{tabular}{lllll}
\tablefirsthead{  
\hline
MJD     &     Vel &   EW & $v\sin i$& note     \\
\hline}
 \begin{supertabular}{llllc}
%\tablefirsthead{  
%\hline
%MJD     &     Vel &   EW & $v\sin i$& note     \\
%\hline\\}
\tablehead{
\hline
MJD     &     Vel &   EW & $v\sin i$& note     \\
%\multicolumn{5}{|l|}{\small\sl continued from previous column}\\
\hline}
\tabletail{
\hline
\multicolumn{5}{|r|}{\small\sl continued on next column}\\
\hline}
\tablelasttail{\hline}
%\bottomcaption{x = o =??}
   51541.28 &    24.9 &  4.5 &21.5  &  o    \\
   51552.29 &    24.5 &  4.0 &23.   &  o    \\
   51553.29 &    23.2 &  3.9 &24.5  &  o    \\
   51585.25 &    23.6 &  4.5 &22.   &  o    \\
   51588.24 &    21.0 &  4.1 &20.5  &  o    \\
   51590.21 &    19.4 &  4.0 &19.   &  o    \\
   51594.25 &    19.1 &  4.0 &18.5  &  o    \\
   51596.10 &    19.2 &  3.9 &17.5  &  o    \\
   51605.11 &    22.6 &  4.4 &17.5  &       \\
   51607.21 &    24.4 &  4.2 &18.   &       \\
   51610.10 &    26.3 &  4.0 &16.   &       \\
   51640.13 &    23.0 &  5.2 &18.5  &       \\
   51641.14 &    23.8 &  5.0 &18.5  &       \\
   51642.06 &    25.2 &  5.0 &19.   &       \\
   51644.14 &    26.9 &  5.2 &18.5  &       \\
   51655.95 &    31.9 &  4.7 &24.5  &       \\
   51658.09 &    30.7 &  4.5 &24.   &       \\
   51661.10 &    30.7 &  4.3 &22.   &       \\
   51664.06 &    28.5 &  4.1 &23.   &       \\
   51672.00 &    29.0 &  4.0 &19.5  &       \\
   51678.02 &    29.4 &  3.6 &18.   &       \\
   51684.06 &    27.9 &  3.7 &18.   &       \\
   51686.90 &    27.2 &  3.8 &20.5  &       \\
   51695.02 &    25.7 &  3.9 &19.   &       \\
   51700.03 &    26.6 &  4.1 &19.   &       \\
   51700.90 &    26.7 &  3.4 &19.5  &       \\
   51702.02 &    26.8 &  4.0 &21.   &       \\
   51706.94 &    25.7 &  3.7 &20.   &       \\
   51708.94 &    25.9 &  3.7 &18.5  &       \\
   51712.94 &    25.0 &  4.0 &20.5  &       \\
   51714.94 &    24.8 &  3.8 &19.5  &       \\
   51740.91 &    21.3 &  4.0 &20.   &       \\
   51741.89 &    21.6 &  3.9 &20.5  &       \\
   51742.89 &    21.2 &  3.8 &18.5  &       \\
   51743.89 &    21.6 &  3.7 &18.5  &       \\
   51744.89 &    22.1 &  3.6 &17.   &       \\
   51746.89 &    22.2 &  4.1 &18.5  &       \\
   51751.94 &    24.2 &  3.8 &18.   &       \\
   51757.87 &    26.3 &  4.0 &20.5  &       \\
   51758.87 &    26.4 &  4.1 &22.5  &       \\
   51760.87 &    25.8 &  4.3 &23.5  &       \\
   51761.86 &    26.2 &  4.0 &24.   &       \\
   51762.86 &    25.7 &  4.1 &23.   &       \\
   51767.85 &    23.8 &  4.4 &23.   &       \\
   51770.84 &    24.6 &  3.9 &20.5  &       \\
   51784.88 &    25.3 &  4.0 &21.5  &       \\
   51785.82 &    24.7 &  3.8 &19.5  &       \\
   51786.82 &    25.2 &  3.8 &19.   &       \\
   51790.89 &    26.0 &  3.7 &20.   &       \\
   51791.84 &    25.5 &  3.8 &20.5  &       \\
   51807.78 &    25.0 &  4.3 &21.   &       \\
   51809.79 &    24.5 &  4.5 &21.5  &       \\
   51810.79 &    25.3 &  4.0 &21.   &       \\
   51811.80 &    25.4 &  4.7 &22.   &       \\
   51813.80 &    25.7 &  4.1 &19.   &       \\
   51815.78 &    26.7 &  4.4 &19.5  &       \\
   51817.81 &    25.1 &  4.5 &20.5  &       \\
   51822.77 &    26.7 &  4.4 &20.5  &       \\
   51830.76 &    29.1 &  4.2 &20.   &       \\
   51836.75 &    29.9 &  3.8 &19.   &       \\
   51847.73 &    26.2 &  9.9 &99.   &       \\
   51849.73 &    26.2 &  4.0 &22.   &       \\
   51851.73 &    25.8 &  4.2 &21.   &       \\
   51860.72 &    28.1 &  4.3 &19.   &       \\
   51861.71 &    28.2 &  4.6 &20.   &       \\
   51864.70 &    28.6 &  4.4 &19.5  &       \\
   51868.27 &    28.9 &  4.0 &21.   &  o    \\
   51870.27 &    30.0 &  3.6 &19.   &  o    \\
   51880.28 &    28.0 &  3.2 &19.5  &  o    \\
   51887.26 &    22.5 &  2.5 &14.5  &  o    \\
   51954.19 &    29.5 &  4.1 &19.   &  o    \\
   51957.22 &    28.5 &  4.0 &21.   &  o    \\
   51967.14 &    24.8 &  4.4 &21.   &       \\
   51970.21 &    24.9 &  4.5 &21.5  &       \\
   51971.21 &    23.3 &  4.0 &20.   &       \\
   51973.23 &    23.0 &  4.2 &21.5  &       \\
   51976.10 &    23.3 &  3.8 &20.   &       \\
   51977.07 &    22.2 &  4.2 &21.   &       \\
   51979.12 &    23.3 &  4.5 &20.   &       \\
   51980.21 &    23.1 &  4.1 &19.   &       \\
   51982.10 &    24.8 &  4.4 &20.5  &       \\
   52028.08 &    22.1 &  4.2 &18.5  &       \\
   52034.09 &    21.9 &  4.3 &18.5  &       \\
   52037.09 &    20.8 &  4.2 &19.   &       \\
   52041.06 &    20.1 &  4.4 &20.   &       \\
   52042.08 &    20.6 &  4.2 &19.   &       \\
   52058.03 &    24.2 &  3.6 &19.5  &       \\
   52068.02 &    25.1 &  4.1 &22.   &       \\
   52069.02 &    24.6 &  4.1 &23.5  &       \\
   52071.00 &    25.1 &  4.3 &23.5  &       \\
   52075.94 &    25.9 &  4.0 &23.5  &       \\
   52078.01 &    23.0 &  4.4 &23.5  &       \\
   52081.05 &    23.9 &  4.0 &21.   &       \\
   52082.99 &    24.2 &  3.7 &21.5  &       \\
   52084.95 &    24.3 &  4.2 &22.5  &       \\
   52088.01 &    24.8 &  3.8 &20.   &       \\
   52089.05 &    24.5 &  4.1 &21.5  &       \\
   52091.00 &    26.1 &  4.4 &21.5  &       \\
   52093.97 &    26.5 &  4.2 &22.   &       \\
   52094.91 &    26.5 &  4.0 &21.5  &       \\
   52097.00 &    25.9 &  3.9 &21.   &       \\
   52098.90 &    26.7 &  4.3 &21.   &       \\
   52100.89 &    28.1 &  3.8 &20.   &       \\
   52105.88 &    28.9 &  4.1 &21.   &       \\
   52113.91 &    27.6 &  4.7 &24.5  &       \\
   52116.87 &    27.4 &  4.6 &23.5  &       \\
   52117.94 &    27.5 &  4.3 &23.5  &       \\
   52119.87 &    28.1 &  4.3 &25.   &       \\
   52122.87 &    30.5 &  4.0 &24.   &       \\
   52126.87 &    29.3 &  9.9 &22.5  &       \\
   52128.86 &    29.4 &  3.7 &22.   &       \\
   52131.89 &    26.7 &  3.7 &22.   &       \\
   52134.85 &    26.4 &  3.6 &21.5  &       \\
   52135.86 &    26.7 &  4.1 &22.5  &       \\
   52137.85 &    26.2 &  3.7 &24.5  &       \\
   52140.90 &    25.6 &  4.0 &23.5  &       \\
   52141.84 &    24.8 &  3.6 &22.5  &       \\
   52145.92 &    21.2 &  3.8 &20.   &       \\
   52174.78 &    26.0 &  3.9 &18.   &       \\
   52181.77 &    25.8 &  3.9 &20.   &       \\
   52185.76 &    26.3 &  4.2 &21.   &       \\
   52188.77 &    25.2 &  3.9 &19.   &       \\
   52200.74 &    23.8 &  4.3 &22.   &       \\
   52206.75 &    21.8 &  3.8 &18.   &       \\
   52207.75 &    21.4 &  3.9 &18.   &       \\
   52212.73 &    23.1 &  4.1 &17.5  &       \\
   52213.73 &    23.1 &  4.3 &19.   &       \\
   52214.73 &    24.1 &  4.1 &16.5  &       \\
   52226.71 &    26.8 &  3.7 &18.5  &       \\
   52255.23 &    20.7 &  3.8 &18.5  &       \\
   52257.29 &    21.4 &  3.9 &19.   &       \\
   52258.29 &    20.6 &  3.9 &19.5  &       \\
   52263.29 &    21.4 &  4.2 &20.5  &       \\
   52265.25 &    21.2 &  4.1 &19.   &       \\
   52275.28 &    22.4 &  3.9 &16.5  &       \\
   52278.29 &    22.3 &  3.8 &18.   &       \\
   52292.25 &    18.4 &  4.1 &21.   &       \\
   52305.13 &    22.1 &  4.4 &19.5  &       \\
   52309.17 &    25.3 &  4.1 &19.5  &       \\
   52319.13 &    24.1 &  4.1 &23.   &       \\
   52326.22 &    21.7 &  4.5 &22.5  &       \\
   52328.21 &    22.4 &  4.1 &23.   &       \\
   52329.18 &    23.2 &  4.0 &20.   &       \\
   52332.19 &    22.3 &  5.1 &23.   &       \\
   52334.18 &    23.6 &  4.0 &19.   &       \\
   52335.17 &    23.4 &  4.3 &19.5  &       \\
   52360.11 &    26.9 &  4.1 &20.5  &       \\
   52361.15 &    26.9 &  4.4 &20.5  &       \\
   52362.11 &    26.2 &  4.1 &20.   &       \\
   52363.15 &    25.6 &  3.9 &19.   &       \\
   52368.16 &    26.6 &  4.0 &19.   &       \\
   52370.11 &    27.5 &  4.2 &21.   &       \\
   52371.11 &    27.3 &  4.2 &22.   &       \\
   52374.03 &    25.5 &  4.0 &24.5  &       \\
   52381.02 &    26.7 &  3.8 &21.5  &       \\
   52382.14 &    26.8 &  3.8 &22.   &       \\
   52384.10 &    27.3 &  3.9 &22.   &       \\
   52388.11 &    28.5 &  3.6 &19.5  &       \\
   52388.99 &    28.5 &  3.7 &20.   &       \\
   52391.08 &    28.2 &  3.9 &18.   &       \\
   52393.98 &    28.3 &  9.9 &99.   &       \\
   52396.08 &    29.3 &  4.1 &19.   &       \\
   52398.10 &    29.6 &  4.4 &19.   &       \\
   52399.00 &    30.4 &  4.0 &18.5  &       \\
   52401.94 &    30.2 &  4.3 &20.   &       \\
   52410.04 &    31.2 &  4.0 &21.5  &       \\
   52411.08 &    30.6 &  9.9 &99.   &       \\
   52421.04 &    27.8 &  4.1 &23.   &       \\
   52422.04 &    27.5 &  4.1 &23.   &       \\
   52423.05 &    27.4 &  3.9 &23.   &       \\
   52423.93 &    25.9 &  3.9 &23.5  &       \\
   52425.03 &    24.4 &  4.0 &25.5  &       \\
   52425.99 &    24.8 &  3.9 &24.   &       \\
   52427.04 &    23.8 &  4.2 &24.   &       \\
   52429.94 &    21.6 &  4.1 &22.5  &       \\
   52434.06 &    20.2 &  4.1 &21.5  &       \\
   52434.93 &    19.9 &  3.9 &20.5  &       \\
   52436.00 &    20.4 &  4.3 &21.5  &       \\
   52440.07 &    21.7 &  4.1 &21.   &       \\
   52442.99 &    22.8 &  4.3 &21.   &       \\
   52443.96 &    22.6 &  4.2 &18.5  &       \\
   52447.92 &    24.1 &  4.4 &19.5  &       \\
   52448.99 &    24.4 &  4.0 &19.   &       \\
   52450.92 &    25.3 &  4.3 &20.5  &       \\
   52451.91 &    25.7 &  4.0 &18.   &       \\
   52452.90 &    25.0 &  4.0 &19.5  &       \\
   52458.97 &    24.4 &  3.8 &20.   &       \\
   52465.01 &    22.0 &  4.0 &21.   &       \\
   52465.90 &    21.1 &  3.8 &20.   &       \\
   52468.90 &    19.6 &  3.7 &19.   &       \\
   52469.90 &    19.4 &  3.8 &19.   &       \\
   52475.98 &    18.2 &  3.9 &19.   &       \\
   52481.88 &    19.5 &  4.2 &19.   &       \\
   52492.87 &    22.5 &  4.2 &20.   &       \\
   52498.85 &    20.8 &  4.1 &20.   &       \\
   52500.85 &    20.6 &  4.2 &20.   &       \\
   52502.85 &    19.9 &  4.1 &20.   &       \\
   52503.86 &    20.0 &  3.8 &18.   &       \\
   52513.84 &    22.0 &  4.1 &19.5  &       \\
   52514.83 &    22.2 &  4.3 &19.   &       \\
   52515.87 &    22.3 &  4.3 &19.5  &       \\
   52518.84 &    23.8 &  4.2 &20.5  &       \\
   52519.86 &    24.7 &  4.3 &21.   &       \\
   52520.83 &    23.9 &  4.4 &20.5  &       \\
   52521.85 &    23.4 &  4.3 &21.   &       \\
   52525.89 &    22.8 &  4.3 &22.   &       \\
   52527.81 &    21.8 &  4.8 &21.   &       \\
   52529.81 &    22.5 &  4.3 &20.   &       \\
   52530.81 &    22.7 &  4.8 &21.   &       \\
   52539.85 &    25.4 &  4.7 &21.5  &       \\
   52540.79 &    26.1 &  4.8 &21.5  &       \\
   52543.79 &    26.2 &  4.1 &21.   &       \\
   52546.81 &    24.8 &  4.3 &22.   &       \\
   52550.77 &    24.6 &  4.0 &20.5  &       \\
   52551.80 &    25.3 &  4.1 &21.   &       \\
   52553.79 &    25.1 &  3.8 &19.5  &       \\
   52559.79 &    22.4 &  4.1 &21.5  &       \\
   52565.73 &    19.6 &  4.0 &20.5  &       \\
   52582.72 &    24.3 &  4.4 &20.   &       \\
   52585.72 &    25.9 &  4.1 &18.   &       \\
   52590.72 &    25.6 &  3.8 &18.5  &       \\
   52591.71 &    25.5 &  4.2 &20.   &       \\
   52613.27 &    24.9 &  4.1 &19.   &       \\
   52617.28 &    24.2 &  3.9 &16.5  &       \\
   52619.22 &    23.3 &  4.2 &19.   &       \\
   52645.21 &    24.2 &  4.0 &17.   &       \\
   52646.20 &    24.9 &  4.2 &18.5  &       \\
   52650.28 &    25.6 &  4.3 &18.5  &       \\
   52655.26 &    26.5 &  3.8 &18.   &       \\
   52666.19 &    26.4 &  4.2 &19.5  &       \\
   52685.27 &    25.9 &  4.0 &17.5  &  o    \\
   52688.21 &    25.9 &  3.1 &18.   &  o    \\
   52689.12 &    23.2 &  2.4 &17.   &  o    \\
   52691.20 &    23.9 &  9.9 &99.   &  o    \\
   52745.08 &    22.2 &  4.2 &19.   &  o    \\
   52748.09 &    24.1 &  4.5 &19.   &  o    \\
   52751.08 &    24.6 &  4.2 &19.   &  o    \\
   52753.05 &    24.6 &  4.2 &19.5  &  o    \\
   52758.09 &    23.4 &  4.3 &21.5  &  o    \\
   52760.10 &    24.3 &  4.3 &20.5  &  o    \\
   52765.05 &    23.7 &  4.5 &22.5  &  o    \\
   52766.99 &    23.4 &  4.2 &22.   &  o    \\
   52769.07 &    23.3 &  4.3 &21.5  &       \\
   52771.07 &    23.1 &  4.3 &21.5  &       \\
   52772.07 &    23.6 &  4.3 &21.   &       \\
   52774.07 &    23.2 &  4.1 &19.   &       \\
   52777.07 &    22.9 &  4.1 &19.   &       \\
   52779.08 &    22.8 &  4.3 &18.5  &       \\
   52783.07 &    23.8 &  4.6 &20.   &       \\
   52785.08 &    24.2 &  4.3 &18.5  &       \\
   52787.08 &    25.0 &  4.5 &19.   &       \\
   52790.99 &    25.9 &  4.1 &19.   &       \\
   52795.99 &    28.6 &  4.1 &19.5  &       \\
   52801.05 &    30.5 &  4.0 &18.5  &       \\
   52803.02 &    31.3 &  4.0 &19.5  &       \\
   52805.05 &    31.7 &  3.9 &20.   &       \\
   52808.05 &    31.1 &  3.9 &22.   &       \\
   52809.04 &    31.4 &  3.6 &20.5  &       \\
   52810.99 &    30.5 &  3.9 &22.5  &       \\
   52814.04 &    27.2 &  3.3 &23.5  &       \\
   52814.97 &    26.8 &  3.8 &25.   &       \\
   52818.04 &    20.6 &  3.4 &21.5  &       \\
   52824.93 &    16.9 &  4.0 &19.   &       \\
   52827.90 &    16.9 &  3.9 &17.   &       \\
   52830.91 &    17.8 &  4.3 &17.5  &       \\
   52832.95 &    17.8 &  4.2 &17.5  &       \\
   52833.94 &    18.1 &  3.9 &16.   &       \\
   52834.95 &    18.2 &  4.2 &17.5  &       \\
   52835.94 &    19.0 &  3.9 &17.   &       \\
   52840.88 &    21.9 &  4.0 &16.5  &       \\
   52841.95 &    21.7 &  4.0 &16.   &       \\
   52847.92 &    25.4 &  4.3 &19.   &       \\
   52853.88 &    28.7 &  3.9 &21.5  &       \\
   52854.88 &    28.2 &  3.9 &23.   &       \\
   52857.88 &    25.4 &  3.6 &24.5  &       \\
   52859.87 &    24.6 &  3.8 &25.5  &       \\
   52860.87 &    24.2 &  4.0 &25.5  &       \\
   52865.91 &    20.5 &  3.5 &22.5  &       \\
   52867.85 &    20.6 &  4.0 &23.   &       \\
   52870.85 &    18.4 &  3.7 &19.5  &       \\
   52871.84 &    18.6 &  4.1 &19.5  &       \\
   52875.83 &    20.6 &  4.1 &19.   &       \\
   52880.82 &    23.2 &  3.8 &17.5  &       \\
   52881.83 &    23.6 &  4.0 &18.   &       \\
   52892.88 &    27.4 &  4.4 &19.   &       \\
   52895.79 &    27.7 &  4.1 &19.   &       \\
   52896.79 &    27.9 &  4.5 &18.5  &       \\
   52899.81 &    27.2 &  4.1 &19.   &       \\
   52901.80 &    26.9 &  4.0 &19.   &       \\
   52905.79 &    27.8 &  3.8 &20.   &       \\
   52911.82 &    23.6 &  3.3 &17.5  &       \\
   52915.78 &    23.0 &  3.7 &20.   &       \\
   52923.79 &    20.6 &  4.0 &18.5  &       \\
   52926.78 &    22.2 &  4.0 &19.   &       \\
   52928.76 &    23.5 &  3.9 &18.   &       \\
   52929.75 &    24.4 &  3.8 &18.   &       \\
   52930.75 &    24.5 &  3.9 &18.   &       \\
   52939.73 &    21.3 &  4.0 &22.   &       \\
   52944.73 &    20.2 &  4.0 &22.   &       \\
   52946.72 &    19.5 &  3.6 &19.5  &       \\
   52947.72 &    21.0 &  3.7 &21.5  &       \\
   52948.72 &    21.7 &  4.3 &22.5  &       \\
   52950.72 &    22.4 &  4.1 &22.   &       \\
   52980.28 &    22.6 &  4.4 &20.   &       \\
   52981.28 &    22.4 &  4.3 &20.   &       \\
   52988.26 &    24.3 &  4.1 &18.5  &       \\
   53000.27 &    27.7 &  4.4 &18.5  &       \\
   53013.24 &    30.0 &  4.5 &23.5  &       \\
   53019.25 &    28.3 &  4.2 &24.   &       \\
   53021.26 &    27.3 &  4.2 &23.5  &       \\
   53034.15 &    27.9 &  4.7 &20.5  &       \\
   53044.20 &    24.0 &  4.1 &22.   &       \\
   53061.20 &    25.0 &  4.3 &19.5  &       \\
   53065.21 &    26.2 &  4.4 &21.   &       \\
   53081.17 &    28.0 &  3.6 &19.   &       \\
   53095.14 &    23.5 &  4.4 &20.5  &       \\
   53098.16 &    24.8 &  3.9 &18.5  &       \\
   53102.08 &    25.8 &  4.4 &18.5  &       \\
   53109.05 &    27.6 &  4.2 &17.   &       \\
   53110.11 &    28.9 &  4.3 &18.   &       \\
   53115.09 &    29.3 &  4.0 &16.5  &       \\
   53117.13 &    29.0 &  4.0 &18.   &       \\
   53119.05 &    28.1 &  4.2 &20.   &       \\
   53131.10 &    22.8 &  3.8 &19.   &       \\
   53132.07 &    22.6 &  3.7 &18.5  &       \\
   53142.08 &    23.6 &  3.8 &17.5  &       \\
   53144.05 &    23.6 &  3.8 &17.   &       \\
   53147.02 &    26.7 &  4.0 &17.5  &       \\
   53149.04 &    26.4 &  3.8 &17.   &       \\
   53156.01 &    27.0 &  3.8 &18.5  &       \\
   53161.03 &    25.0 &  3.9 &23.   &       \\
   53164.02 &    23.9 &  3.9 &22.   &       \\
   53169.01 &    22.3 &  3.7 &22.5  &       \\
   53173.00 &    19.4 &  3.7 &21.   &       \\
   53178.00 &    18.3 &  3.7 &19.   &       \\
   53181.99 &    19.5 &  3.9 &18.5  &       \\
   53183.90 &    20.3 &  3.9 &18.5  &       \\
   53188.99 &    22.3 &  4.0 &17.5  &       \\
   53191.97 &    23.7 &  4.1 &18.5  &       \\
   53195.97 &    24.5 &  4.0 &18.   &       \\
   53202.89 &    23.3 &  4.1 &19.   &       \\
   53212.89 &    20.2 &  3.8 &20.5  &       \\
   53216.87 &    20.8 &  3.7 &17.   &       \\
   53224.86 &    25.2 &  3.9 &18.   &       \\
   53227.97 &    25.1 &  3.7 &18.5  &       \\
   53236.91 &    25.1 &  3.8 &21.5  &       \\
   53246.82 &    22.9 &  4.0 &21.   &       \\
   53248.93 &    21.5 &  3.9 &21.5  &       \\
   53249.85 &    20.9 &  3.7 &21.   &       \\
   53253.86 &    19.3 &  3.7 &19.5  &       \\
   53255.81 &    19.8 &  3.8 &19.5  &       \\
   53261.80 &    21.4 &  4.0 &17.5  &       \\
   53263.79 &    22.2 &  4.1 &17.   &       \\
   53266.80 &    24.0 &  4.1 &18.   &       \\
   53283.77 &    21.3 &  3.2 &20.   &       \\
   53303.73 &    23.1 &  4.0 &19.5  &       \\
   53313.71 &    26.9 &  3.7 &20.   &       \\
   53322.70 &    23.9 &  3.8 &24.   &       \\
   53323.70 &    22.5 &  4.0 &26.   &       \\
   53328.70 &    21.6 &  4.4 &25.   &       \\
   53357.24 &    28.7 &  4.4 &21.5  &       \\
   53365.29 &    26.1 &  4.6 &21.5  &       \\
   53379.28 &    28.4 &  4.4 &22.   &       \\
   53383.26 &    26.2 &  4.1 &23.   &       \\
   53392.27 &    23.1 &  3.8 &20.   &       \\
   53441.20 &    26.8 &  3.8 &17.   &       \\
   53452.19 &    28.2 &  4.0 &21.   &       \\
   53479.11 &    22.1 &  4.0 &16.   &       \\
   53482.07 &    23.9 &  4.0 &15.   &       \\
   53495.05 &    28.8 &  4.4 &18.   &       \\
   53498.05 &    29.4 &  3.9 &18.5  &       \\
   53501.08 &    29.4 &  3.9 &20.   &       \\
   53502.06 &    28.8 &  4.0 &20.5  &       \\
   53505.05 &    26.2 &  3.7 &21.   &       \\
   53513.06 &    20.8 &  3.7 &16.5  &       \\
   53518.00 &    19.8 &  3.8 &17.   &       \\
   53521.00 &    18.8 &  3.9 &17.5  &       \\
   53522.02 &    19.8 &  3.6 &17.   &       \\
   53527.98 &    20.6 &  3.9 &17.   &       \\
   53529.91 &    21.8 &  4.1 &17.5  &       \\
   53532.02 &    23.1 &  4.1 &18.5  &       \\
   53534.94 &    25.1 &  3.9 &19.   &       \\
   53542.91 &    28.2 &  3.9 &21.   &       \\
   53547.93 &    27.5 &  3.5 &21.   &       \\
   53560.88 &    24.6 &  3.8 &23.   &       \\
   53567.88 &    23.7 &  3.6 &21.5  &       \\
   53568.89 &    23.9 &  3.8 &22.   &       \\
   53571.93 &    22.6 &  4.0 &20.5  &       \\
   53579.87 &    25.6 &  3.7 &18.   &       \\
   53584.87 &    29.7 &  4.1 &18.5  &       \\
   53588.87 &    30.5 &  4.1 &20.   &       \\
   53590.91 &    30.3 &  4.0 &20.   &       \\
   53597.85 &    32.2 &  3.7 &19.5  &       \\
   53598.84 &    31.8 &  3.4 &19.5  &       \\
   53599.85 &    32.0 &  3.6 &20.5  &       \\
   53603.84 &    30.3 &  3.3 &23.   &       \\
   53606.88 &    27.4 &  4.1 &99.   &       \\
   53607.84 &    24.9 &  3.7 &26.5  &       \\
   53609.85 &    23.6 &  3.8 &25.5  &       \\
   53612.82 &    23.0 &  3.9 &24.5  &       \\
   53615.87 &    22.1 &  3.7 &23.   &       \\
   53620.81 &    21.5 &  4.0 &20.5  &       \\
   53621.83 &    20.7 &  3.7 &19.   &       \\
   53625.80 &    22.4 &  4.0 &18.   &       \\
   53627.79 &    23.4 &  3.9 &18.   &       \\
   53629.83 &    24.6 &  4.2 &20.   &       \\
   53636.80 &    26.1 &  4.0 &19.   &       \\
   53638.78 &    27.5 &  3.5 &18.5  &       \\
   53645.77 &    26.3 &  3.9 &25.   &       \\
   53647.77 &    24.2 &  3.7 &24.5  &       \\
   53658.76 &    19.5 &  4.1 &20.5  &       \\
   53663.74 &    20.0 &  3.5 &18.5  &       \\
   53668.74 &    21.1 &  3.9 &17.5  &       \\
   53670.73 &    20.5 &  3.8 &18.   &       \\
   53672.72 &    21.2 &  3.8 &18.   &       \\
   53674.72 &    21.8 &  4.0 &19.   &       \\
   53686.71 &    23.1 &  3.9 &22.5  &       \\
   53690.71 &    21.3 &  4.2 &22.   &       \\
   53692.70 &    21.4 &  4.2 &21.   &       \\
   53721.28 &    18.6 &  4.4 &21.5  &       \\
   53764.28 &    26.4 &  3.6 &21.5  &       \\
   53782.15 &    27.0 &  3.5 &18.   &       \\
   53795.19 &    28.0 &  4.4 &22.   &       \\
   53800.16 &    26.9 &  4.3 &23.5  &       \\
   53817.16 &    26.8 &  4.1 &21.5  &       \\
   53829.12 &    24.1 &  4.0 &20.5  &       \\
   53831.10 &    25.2 &  4.1 &19.   &       \\
   53834.09 &    26.8 &  4.3 &21.5  &       \\
   53836.10 &    26.4 &  4.2 &23.   &       \\
   53851.06 &    24.0 &  4.2 &26.   &       \\
   53857.04 &    24.3 &  4.4 &23.5  &       \\
   53861.11 &    24.9 &  4.3 &21.5  &       \\
   53864.04 &    24.0 &  4.2 &23.   &       \\
   53871.04 &    24.6 &  4.0 &21.   &       \\
   53881.99 &    23.8 &  3.9 &18.5  &       \\
   53885.01 &    22.9 &  3.9 &19.5  &       \\
   53888.98 &    23.1 &  4.4 &20.   &       \\
   53894.96 &    25.2 &  4.6 &20.5  &       \\
   53898.01 &    27.0 &  4.4 &21.   &       \\
   53907.98 &    26.5 &  4.3 &25.5  &       \\
   53909.01 &    27.0 &  4.2 &25.   &     x \\
   53910.98 &    25.8 &  4.1 &24.5  &     x \\
   53913.97 &    24.8 &  4.3 &26.   &     x \\
   53914.97 &    24.7 &  4.3 &26.5  &     x \\
   53916.96 &    23.6 &  4.1 &24.5  &     x \\
   53918.97 &    22.9 &  4.1 &23.5  &     x \\
   53919.96 &    23.4 &  4.1 &24.5  &     x \\
   53924.04 &    22.3 &  3.8 &21.   &     x \\
   53925.90 &    22.1 &  4.0 &20.   &     x \\
   53927.90 &    22.6 &  3.9 &19.5  &     x \\
   53928.90 &    23.0 &  4.2 &20.   &     x \\
   53929.89 &    23.4 &  3.9 &19.   &     x \\
   53930.94 &    23.2 &  4.1 &19.   &     x \\
   53931.91 &    23.7 &  4.3 &19.5  &     x \\
   53932.89 &    23.7 &  4.2 &19.5  &     x \\
   53933.90 &    24.3 &  4.2 &19.5  &     x \\
   53934.90 &    24.8 &  4.1 &19.   &     x \\
   53936.90 &    25.8 &  4.3 &21.   &     x \\
   53939.96 &    26.0 &  4.2 &20.5  &     x \\
   53940.91 &    25.8 &  4.2 &20.5  &     x \\
   53944.99 &    25.6 &  4.1 &20.5  &     x \\
   53948.98 &    24.3 &  4.0 &21.   &     x \\
   53953.87 &    21.4 &  3.8 &22.5  &     x \\
   53954.90 &    22.0 &  3.8 &22.   &     x \\
   53956.90 &    21.3 &  4.0 &21.5  &     x \\
   53960.85 &    21.0 &  3.7 &20.   &     x \\
   53962.93 &    20.3 &  3.7 &19.5  &       \\
   53975.84 &    24.6 &  4.5 &22.   &       \\
   53976.83 &    25.6 &  4.4 &21.5  &       \\
   53985.87 &    23.5 &  3.8 &20.5  &       \\
   53986.82 &    22.4 &  4.1 &21.5  &       \\
   53998.82 &    21.4 &  3.6 &20.5  &       \\
   54000.80 &    22.3 &  3.6 &20.   &       \\
   54032.74 &    21.8 &  3.9 &20.5  &       \\
   54034.73 &    21.9 &  3.9 &19.   &       \\
   54040.73 &    23.0 &  3.7 &18.   &       \\
   54042.72 &    22.8 &  3.9 &18.5  &       \\
   54045.71 &    22.8 &  3.3 &16.5  &       \\
   54048.72 &    22.2 &  3.7 &18.   &       \\
   54057.70 &    19.7 &  3.6 &19.5  &       \\
   54065.27 &    20.1 &  4.0 &22.   &       \\
   54078.28 &    26.7 &  4.6 &21.5  &       \\
   54086.29 &    27.9 &  4.6 &22.   &       \\
   54113.24 &    26.6 &  4.4 &23.5  &       \\
   54134.28 &    20.5 &  4.3 &21.5  &       \\
   54146.26 &    26.1 &  4.8 &22.5  &       \\
   54161.23 &    26.5 &  4.2 &21.5  &       \\
   54181.17 &    21.5 &  4.6 &17.   &       \\
   54185.00 &    23.9 &  4.3 &17.5  &       \\
   54192.13 &    24.6 &  4.2 &21.   &       \\
   54194.13 &    24.2 &  4.3 &21.5  &       \\
   54200.10 &    22.6 &  4.7 &25.   &       \\
   54206.11 &    20.4 &  4.8 &23.5  &       \\
   54220.09 &    21.6 &  5.1 &22.   &       \\
   54228.06 &    23.3 &  5.5 &24.5  &       \\
   54235.06 &    23.8 &  5.1 &26.   &       \\
   54239.05 &    25.1 &  5.1 &27.   &       \\
   54243.07 &    24.8 &  5.0 &27.   &       \\
   54250.04 &    24.0 &  4.7 &23.   &       \\
   54251.01 &    23.7 &  4.8 &23.   &       \\
   54255.99 &    22.2 &  4.4 &23.   &       \\
   54267.00 &    21.9 &  4.2 &19.5  &       \\
   54271.98 &    23.3 &  4.5 &21.   &       \\
   54277.99 &    24.6 &  4.6 &23.   &       \\
   54287.99 &    20.9 &  5.8 &34.   &     o \\
   54288.98 &    22.2 &  5.3 &32.5  &     o \\
   54293.99 &    19.5 &  5.7 &36.   &     o \\
   54299.90 &    19.1 &  6.1 &37.   &     o \\
   54300.92 &    20.3 &  5.6 &36.   &     o \\
   54303.90 &    20.8 &  5.3 &33.   &     o \\
   54307.99 &    20.8 &  4.1 &29.   &     o \\
   54310.92 &    18.1 &  2.0 &27.5  &     o \\
%\end{tabular}
\end{supertabular}
\end{center}

\end{document}